\documentclass[reprint,aps,pre,showpacs]{revtex4-1}

\usepackage{graphicx}
\usepackage{bm}
\usepackage{hyperref}
\usepackage[latin1]{inputenc}
\usepackage{amsmath}
\usepackage{amsfonts}
\usepackage{amssymb}
\usepackage{amsthm}

\makeatletter
\newcommand{\colorcaption}[2][]{%
  \begingroup%
  \renewcommand{\@caption@fignum@sep}{ (Color online). }%
  \caption[#1]{#2}%
  \endgroup%
}
\makeatother

\begin{document}


\title{From dynamical systems with time-varying delay to circle maps and Koopmanism}

\author{David M\"uller}
 \email{david.mueller@physik.tu-chemnitz.de}
\author{Andreas Otto}%
 \email{andreas.otto@physik.tu-chemnitz.de}
\author{G\"unter Radons}
 \email{radons@physik.tu-chemnitz.de}
\affiliation{Institute of Physics, Chemnitz University of Technology, 09107 Chemnitz, Germany}
%


\date{\today}

\begin{abstract}
In the present paper we investigate the influence of the retarded access by a time-varying delay on the dynamics of delay systems. We show that there are two universality classes of delays, which lead to fundamental differences in dynamical quantities such as the Lyapunov spectrum. Therefore we introduce an operator theoretic framework, where the solution operator of the delay system is decomposed into the Koopman operator describing the delay access and an operator similar to the solution operator known from systems with constant delay. The Koopman operator corresponds to an iterated map, called access map, which is defined by the iteration of the delayed argument of the delay equation. The dynamics of this one-dimensional iterated map determines the universality classes of the infinite-dimensional state dynamics governed by the delay differential equation. In this way, we connect the theory of time-delay systems with the theory of circle maps and the framework of the Koopman operator.
In the present paper we extend our previous work [Otto, M\"uller, and Radons, Phys. Rev. Lett. 118, 044104 (2017)], by elaborating the mathematical details and presenting further results also on the Lyapunov vectors.
\end{abstract}

\maketitle


\section{\label{sec:intro} Introduction}

Time delay systems emerge, for example, in control theory \citep{2003richard, *2007michiels}, engineering \citep{2010kyrychko, *2011insperger}, life science \citep{1993kuang,1992gopolsamy, *2009Stepan, *2011smith}, chaos control \citep{1992pyragas, *2007schuster} and synchronization of networks \citep{2004atay, *2011lakshmanan}.  In all of these fields effects of time-varying delays are investigated. Introducing time-varying delays can improve the stability of machining processes \citep{2008zatarain,*2011otto} and of systems in general via time-delayed feedback control \citep{2010gjurchinovski, *2010konishi}. Furthermore, in biological models time-varying delays arise naturally by taking into account fluctuations of the environment \citep{1993kuang,2000schley} and also the effect on synchronization is studied \citep{2004kye, *2009ambika}. The functional structure of the delay itself influences mathematical properties of general delay systems such as causality, minimality \cite{2011verriest} and spectral reachability \cite{2012verriest}.

For some systems with time-varying delay it is known that they can be transformed to systems with constant delay because the delay is defined by an intrinsic constant delay and a known timescale transformation \citep{1968seddon,*1993tsao}. This type of systems arises, for example, in biological models with \emph{threshold delays}, where the intrinsic constant delay represents the evolutionary steps, which have to be passed to reach the state of adulthood and the corresponding timescale represents the grade of evolution \cite{1993kuang,1993smith}. Further examples are systems with \emph{variable transport delays}, where the constant delay is given by the length of the transport line and the corresponding timescale represents the covered distance \cite{otto_transformations_2017}. The question arises whether one can find a transformation for a general system with time-varying delay, such that the initial system with time-varying delay is equivalent to a resulting system with constant delay, or not. In other words, is there a well-behaved transformation in the sense that dynamical quantities are well-defined for the new system with constant delay? And are they invariant under the transformation and does the new system correspond to a reasonable physical model, characterized by transport or fixed evolutionary steps? In general, this is not true and the existence of such a well-behaved transformation depends on the functional properties of the time-varying delay. In fact, the resulting delay classification implies fundamental differences in the dynamics of systems that are equivalent to systems with constant delay and the remaining systems, as demonstrated in \cite{2016otto}.

So let us consider a general $d$-dimensional dynamical system defined by the system of delay differential equations (DDE) with time-varying delay $\tau(t)$  
\begin{equation}
\label{eqn:sys1}
\dot{\bm{z}}(t)=\bm{f}(\bm{z}(t),\bm{z}(R(t)),t) \text{, where } R(t):=t-\tau(t)
\end{equation}
is called \emph{retarded argument}. Hence, the results presented below hold for all DDEs with one time-varying delay $\tau(t)$. In other words, the results are independent of the specific form of $\bm{f}$ and apply to autonomous and non-autonomous systems. Although the present theory holds only for systems with one delay, it provides a basis for a more general theory including systems with multiple delays and, even more general, distributed delays.

In this paper we focus on the influence of the functional structure of the retarded argument on the state dynamics as given by Eq.~\eqref{eqn:sys1} and extend the theory established in \cite{2016otto}. We will show that there is a natural connection between the dynamics of systems with time-varying delay, the  dynamics of one-dimensional iterated maps and the spectral properties of the Koopman operator associated with these maps. This connection itself is interesting because it relates well understood and distinct theories and may lead to a better understanding of delay systems with time-varying delay. The analysis of dynamical systems in the framework of the Koopman operator is a wide topic of past and recent research \cite{1931koopman,2012budisic} and is applied, for example, in the computation of isostables and isochrons \cite{2012mauroy,2013mauroy} or in global stability analysis \cite{mauroy_global_2016}.

The paper is organized as follows. In Sec.~\ref{sec:class} we derive the two delay classes as sketched in \cite{2016otto} and give a short overview on the differences in the dynamics. Sec.~\ref{sec:prelim} and Sec.~\ref{sec:finsec} deal with the separation of the dynamics related to the variable delay from the dynamics of the delay system. The influence of the delay classes on the Lyapunov spectrum is analyzed in Sec.~\ref{sec:specdelayclasses} and on the Lyapunov vectors in Sec.~\ref{sec:vec}, thereby explaining also the differences in the numerical properties of spectral methods.

\section{\label{sec:class} Delay classification}

We showed in \cite{2016otto} that two different types of delays can be identified, which result in fundamental differences in the dynamics of the corresponding delay system. In the following we briefly introduce this classification by the analysis of the existence of a timescale transformation that transforms the system of Eq.~\eqref{eqn:sys1} to a system with constant delay. Furthermore, we elaborate the differences in the dynamics and numerics of systems belonging to one or the other delay type.

Every system of differential equations is equivalent to a large set of systems of similar type that are connected among each other by timescale transformations. Such a transformation is easily done by the substitution of the original timescale $t$ by an invertible and at least once differentiable function $t=\Phi(t')$ of the new timescale $t'$. In other words one introduces the new variable $\bm{y}(t')$ by
\begin{equation}
\label{eqn:newvar}
\bm{y}=\bm{z} \circ \Phi,
\end{equation}
where $\circ$ denotes function composition and $\bm{z}(t)$ is the original variable. By the substitution of the new variable $\bm{y}(t')$ into the original delay Eq.~\eqref{eqn:sys1} and by some rearrangements we obtain an equivalent delay system \cite{1965bellman}
\begin{equation}
\label{eqn:sys2}
\dot{\bm{y}}(t')=\dot{\Phi}(t') \cdot \bm{f}(\bm{y}(t'),\bm{y}(R_c(t')),\Phi(t')),
\end{equation}
where the new retarded argument $R_c(t')$ is derived from the original retarded argument by
\begin{equation}
\label{eqn:conju1}
R_c=\Phi^{-1} \circ R \circ \Phi.
\end{equation}
If we find at least one function $\Phi(t')$ such that
\begin{equation}
\label{eqn:retargconst}
R_c(t')=t'-c, \text{ with } c > 0,
\end{equation}
we are able to transform the system with time-varying delay, defined by Eq.~\eqref{eqn:sys1}, into a system with constant delay, Eq.~\eqref{eqn:sys2}. Equation~\eqref{eqn:conju1} takes the structure of a conjugacy equation, which is known from the theory of one-dimensional iterated maps \cite{1997katok}. From this point of view such an equation determines the topological conjugacy between the iterated maps
\begin{eqnarray}
\theta_{k} &=& R(\theta_{k-1}) = R^k(\theta_0)\quad\text{and} \label{eqn:accessmap}\\
\theta'_{k} &=& R_c(\theta'_{k-1})=R_c^k(\theta'_0).
\end{eqnarray}
Since these maps describe the dynamics of the access to the delayed state of our delay system, we call them \emph{access maps}.

In the following we only consider invertible retarded arguments $R(t)$, i.e. $\dot{\tau}(t) < 1$, because non-invertible retarded arguments cause problems with minimality and causality, and are even considered unphysical \cite{2011verriest}.
Now let us assume further that the delay $\tau(t)$ is periodic with period $1$, i.e., we measure time in units of the delay period. Since the related maps can be reduced to maps acting on the interval $[0,1]$ we are able to apply the theory of circle maps \cite{1997katok} and it follows that for some retarded arguments $R(t)$ the corresponding access map is topological conjugate to a pure rotation. The topological conjugacy between one-dimensional iterated maps acting on the same interval preserves the rotation number, which is defined by \cite{1997katok}
\begin{equation}
\label{eqn:rotnr}
\rho = \lim\limits_{k\to\infty} \frac{R^k(t)-t}{k} = -c.
\end{equation}
If the retarded argument $R(t)$ defines a map that is topological conjugate to a pure rotation, the constant delay $c$ in Eq.~\eqref{eqn:retargconst} is determined by Eq.~\eqref{eqn:rotnr} and it follows directly from Eq.~\eqref{eqn:conju1} that $R(t)$ fulfills the criterion
\begin{equation}
\label{eqn:crit}
\lim\limits_{j\to\infty} R^{q_j}(t)+p_j=t \quad \forall t,
\end{equation}
where $\frac{p_j}{q_j}$ is the $j$-th convergent of the continued fraction expansion \cite{khinchin_continued_1997} of $c=-\rho$. Circle maps fulfilling this criterion exhibit marginally stable quasi-periodic (periodic) motion since they are topological conjugate to the pure rotation by an irrational (rational) angle $c$. Invertible maps that do not fulfill the criterion above have stable fixed points or periodic orbits. This follows directly from Poincare's classification \cite{1997katok}.

\begin{figure}
\includegraphics[width=0.45\textwidth]{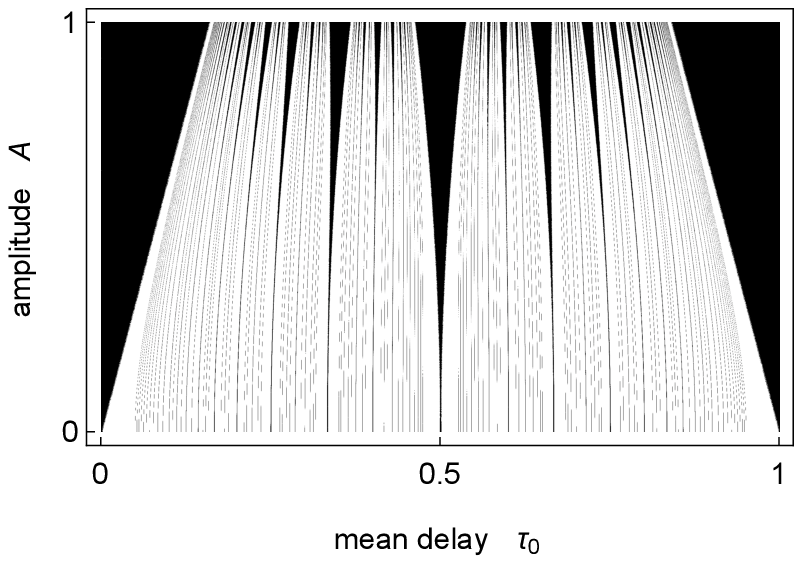}\\
\colorcaption{\label{fig:arnold} Parameter space of an arbitrary system with time-varying delay $\tau(t)=\tau_0+\frac{A}{2\pi} \sin(2\pi\,t)$. The white regions correspond to a \emph{conservative} delay, where the system is equivalent to a system with constant delay. On the other hand, the black regions correspond to a \emph{dissipative} delay, where the system is not equivalent to a system with constant delay.}
\end{figure}

Thus, the question whether a system of DDEs is equivalent to a system with constant delay can be reduced to the question of the existence of a topological conjugacy between the access map defined by the retarded argument $R(t)$ and a pure rotation. Equivalence means that the systems exhibit identical dynamics and that characteristic quantities are invariant under the timescale transformation \footnote{By the iterative method given in \cite{2009brunner} a timescale transformation to constant delay can be computed even in the case, where the access map is not conjugate to a pure rotation. However, in this case the transformation is well-defined only on finite time intervals but not for infinite times. Hence, asymptotic quantities such as Lyapunov exponents are not well-defined for the resulting systems.}. The connection to the existence of a circle map conjugacy leads to interesting consequences for systems with parameter families of delays. As illustrated in Fig.~\ref{fig:arnold}, the parameter regions that are related to a system that is equivalent to a system with constant delay form a Cantor set, whereas regions related to systems that are not equivalent to a system with constant delay are characterized by Arnold tongues.

Obviously the classification depends only on the properties of the access map and not on the specific DDE. In other words, the classification is independent of the specific properties of $\bm{f}$. So if there is any influence of the delay type on the systems dynamics, the influence emerges in all types of time-delay systems and represents a universal property of all systems with time varying delay.

\begin{figure}
\raggedright
\hspace{5mm} a)\\
\centering
\includegraphics[width=0.4\textwidth]{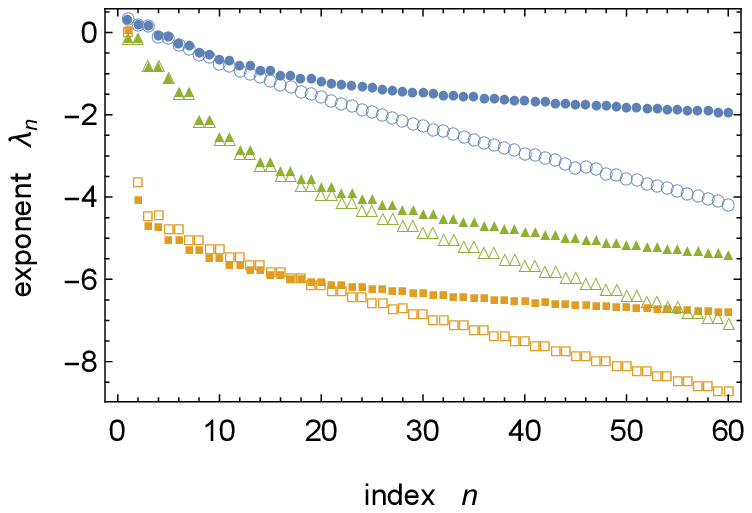}\\
\raggedright
\hspace{5mm} b)\\
\centering
\includegraphics[width=0.4\textwidth]{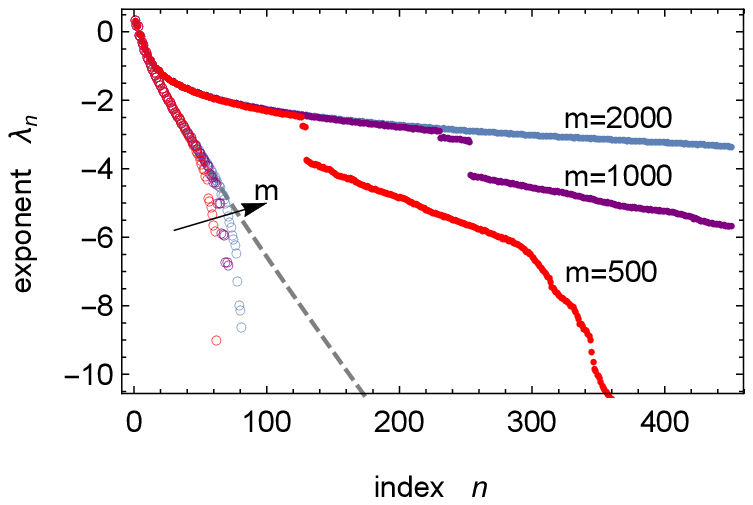}
\colorcaption{\label{fig:spectra} Comparison of the Lyapunov spectra between systems with conservative delay (filled) and dissipative delay (empty), computed by the \emph{semi-discretization} method \cite{2004insperger,2011insperger}. a) The Lyapunov spectra are shown for three different systems, where the systems from top to bottom are given by Eq.~\eqref{eqn:mackey} (circles), Eq.~\eqref{eqn:drei} (triangles) and Eq.~\eqref{eqn:hutch} (squares). b) The convergence properties of the method are illustrated for Eq.~\eqref{eqn:mackey}, where $m$ denotes the number of equidistant points approximating the state of the delay system. As a guide to the eye the asymptotic scaling behavior for dissipative delay, which we derive in Sec.~\ref{sec:specdelayclasses}, is represented by the dashed gray line. The jumps in the spectra related to conservative delay and the deviation of the spectra from the dashed line for dissipative delay, respectively, indicate the bounds on the number of well approximated exponents. For all computations we have chosen the delay $\tau(t)=\tau_0+0.9/(2\pi)\sin(2\pi\,t)$ with the mean delay $\tau_0=1.51$ and $\tau_0=1.54$ for dissipative and conservative delay, respectively.}
\end{figure}

The influence of the delay type on the systems dynamics and its universality can be demonstrated easily by the evolution of small perturbations and the related relaxation rates called \emph{Lyapunov exponents}. The set of all Lyapunov exponents ordered from the largest to the smallest is called \emph{Lyapunov spectrum}. As already argued in \cite{2016otto} the asymptotic scaling behavior of the Lyapunov spectrum depends on the delay class. In Fig.~\ref{fig:spectra} exemplary Lyapunov spectra of three delay systems are shown for each delay type, whereby the spectra were computed directly from the systems with time-varying delay without applying any transformation. The systems are given by
\begin{subequations}
  \begin{align}
\dot{z}(t) &=2 z(t) \left(1-z(R(t))\right), \label{eqn:hutch} \\
\dot{z}(t) &=\frac{10 z(R(t))}{1+z(R(t))^{10}} - 5 z(t), \label{eqn:mackey} \\
\ddot{z}(t) &+ \dot{z}(t) + 4 \pi^2 z(t) = 8 z(R(t)), \label{eqn:drei}
  \end{align}
\end{subequations}
where for constant delay Eq.~\eqref{eqn:hutch} is the Hutchinson equation arising from population dynamics \cite{1992gopolsamy}, Eq.~\eqref{eqn:mackey} is the Mackey-Glass equation known as a model for blood-production \cite{1977mackey} and Eq.~\eqref{eqn:drei} arises in the stability analysis of turning processes \cite{2008zatarain}. To illustrate the universality of the following results, i.e., the independence of the specific dynamics of the system, the parameters and the time-varying delay are chosen such that Eq.~\eqref{eqn:hutch} exhibits a limit cycle, Eq.~\eqref{eqn:mackey} chaotic dynamics, and Eq.~\eqref{eqn:drei} a stable fixed point, respectively.

If the access map of a time-delay system is topological conjugate to a pure rotation, the dynamics of the system equals the dynamics of a system with constant delay. Since the asymptotic scaling behavior of the Lyapunov spectrum for systems with constant delay is known to be logarithmic with respect to the index of the Lyapunov exponents \cite{1982farmer}, the scaling behavior of equivalent systems with time-varying delay is logarithmic as well. We call the delays with this property \emph{conservative delay}, because the related access map is a measure preserving system and has no influence on the scaling behavior of the Lyapunov spectrum. If the access map does not fulfill the criterion Eq.~\eqref{eqn:crit}, the specific delay system is not equivalent to a system with constant delay and consequently, the asymptotic scaling behavior is not logarithmic as is shown in Fig.~\ref{fig:spectra}a. Since the Lyapunov spectrum decreases asymptotically faster than any logarithm we call the corresponding delays \emph{dissipative delay}. Later we will derive in detail that the scaling behavior is asymptotically linear, which was already briefly demonstrated in \cite{2016otto}. The delay type also influences the convergence properties of numerical algorithms as illustrated for the Mackey-Glass equation in Fig.~\ref{fig:spectra}b and for the remaining systems in Fig.~\ref{fig:spectra_app} in Appendix~\ref{sec:applyaspec}. For conservative delays the number of well approximated exponents increases much faster by increasing the number $m$ of sampling points than for disspative delay.

We see that the classification by the access map's dynamics leads to interesting hitherto unknown phenomena such as the above mentioned influence on the Lyapunov spectra and the numerical properties of their computation. We will discuss this in more detail in Sec.\ref{sec:vec}. In the next section we separate the influence of the access map's dynamics from the dynamics of the delay system using a specific operator framework.

\section{\label{sec:prelim} Delay Dynamics in operator framework}

In this section we give some preliminaries for the separation of the influence of the access map on the dynamics of the delay system. The method is derived from the \emph{method of steps} for systems with time-varying delay, cf. \cite{1965bellman}. The separation will be used in Sec.~\ref{sec:finsec} and Sec.~\ref{sec:specdelayclasses} for the explanation of the different asymptotic scaling behavior of the Lyapunov spectra in Fig.~\ref{fig:spectra}.

For a given solution $\tilde{\bm{z}}(t)$ of Eq.~\eqref{eqn:sys1}, the linearized system for the dynamics of small perturbations $\bm{x}(t)=\bm{z}(t)-\tilde{\bm{z}}(t)$ is given by
\begin{equation}
\label{eqn:linsys}
\dot{\bm{x}}(t)=\bm{A}(t)\,\bm{x}(t) + \bm{B}(t)\,\bm{x}(R(t)),
\end{equation}
where $\bm{A}(t)=\partial_{\bm{z}(t)} \bm{f}$ and $\bm{B}(t)=\partial_{\bm{z}(R(t))} \bm{f}$ denote the Jacobians of $\bm{f}$ with respect to the variable $\bm{z}(t)$ and the delayed variable $\bm{z}(R(t))$ taken at the unperturbed solution $\tilde{\bm{z}}(t)$. Consequently, without any loss of generality we can restrict our analysis to $d$-dimensional linear delay systems.

In order to apply the method of steps to our delay system it is convenient to remove the first term on the right hand side of Eq.~\eqref{eqn:linsys} to reduce the computation of the solution to a simple weighted integration of the systems history. Hence we use the variation of constants approach and obtain the integral formulation of Eq.~\eqref{eqn:linsys} 
\begin{equation}
\label{eqn:intform}
\bm{x}(t) = \bm{M}(t,t_0)\,\bm{x}(t_0) + \int_{t_0}^{t} dt'\, \bm{M}(t,t')\,\bm{B}(t')\,\bm{x}(R(t')),
\end{equation}
where $t_0$ denotes the initial time and $\bm{M}(t,t')$ denotes the fundamental solution of the ODE-part and solves
\begin{equation}
\frac{d}{dt}\bm{M}(t,t')=\bm{A}(t)\,\bm{M}(t,t'), \quad \bm{M}(t',t')=\bm{E}_d, 
\end{equation}
where $\bm{E}_d$ is the $d$-dimensional identity matrix. The operator on the right hand side of Eq.~\eqref{eqn:intform} can be decomposed into the operators $\hat{I}$ and $\hat{C}$ and the integral formulation of our delay equation can be represented by the fixed point equation
\begin{equation}
\label{eqn:fixdde}
\bm{x}(t)=\hat{I}[\hat{C}[\bm{x}(t''),t'],t],
\end{equation}
where $\hat{C}$ denotes the \emph{Koopman operator} also called \emph{composition operator} with the symbol $R(t)$
\begin{equation}
\label{eqn:koopmanopdef}
\hat{C}[\bm{x}(t'),t] = \bm{x}(R(t))
\end{equation}
and $\hat{I}$ denotes the integral operator
\begin{eqnarray}
&\hat{I}[\bm{x}(t'),t]& \\ = &\bm{M}(t,t_0) &\,\bm{x}(R^{-1}(t_0)) + \int_{t_0}^{t} dt'\, \bm{M}(t,t')\,\bm{B}(t')\,\bm{x}(t'). \nonumber
\end{eqnarray}

The main idea of the method of steps is to divide the timescale into several subintervals, such that the solution of the DDE reduces to the solution of an inhomogeneous ODE. Trivially the intervals are chosen properly if the interval boundaries $t_k$ are connected by
\begin{equation}
\label{eqn:intbord}
t_{k-1}=R(t_{k}),
\end{equation}
thus giving further meaning to the access map, Eq.~\eqref{eqn:accessmap}.
Hence, the temporal evolution of a DDE can be described by the iteration of operators acting on functions, which are defined on intervals ordered in time. To be consistent with our preceding definitions we define the initial state $\bm{x}_0(t)$ as the initial function inside the interval $[R(t_0),t_0]$ and the state $\bm{x}_k$ in the time-interval after $k$ time-steps as the function
\begin{equation}
\bm{x}_k(t) = \bm{x}(t), \quad t \in [t_{k-1}, t_{k}].
\end{equation}

To adapt the decomposition of the solution operator of our DDE to the aforementioned notation we introduce the restrictions $\hat{I}_k$ and $\hat{C}_k$ of the integration and the Koopman operator, respectively, related to the previously defined time-intervals. The operator $\hat{C}_k$ is defined by the restriction of the Koopman operator $\hat{C}$ to the domain of the related state $\bm{x}_{k}(t)$ living in an appropriate space $\mathcal{F}\left([t_{k-1}, t_{k}],\mathbb{R}^d\right)$ of functions mapping $[t_{k-1}, t_{k}]$ to $\mathbb{R}^d$. We call the family of operators $\hat{C}_k$ \emph{access operators} because they describe the access of the delay system to the preceding state.
\begin{eqnarray}
\label{eqn:compo2}
&& \hat{C}_k : \mathcal{F}\left([t_{k-1}, t_{k}],\mathbb{R}^d\right) \rightarrow \mathcal{F}\left([t_{k}, t_{k+1}],\mathbb{R}^d\right) :  \nonumber\\
&& \hat{C}_k[\bm{x}_{k}(t'),t] = \bm{x}_{k}(R(t)), \quad t \in [t_{k}, t_{k+1}]
\end{eqnarray}
The restriction $\hat{I}_k$ of the integration operator $\hat{I}$ in our notation of labeled time-intervals acts on the space $\mathcal{F}\left([t_{k}, t_{k+1}],\mathbb{R}^d\right)$ where our future state $\bm{x}_{k+1}(t)$ lives and takes the form
\begin{widetext}
\begin{equation}
\label{eqn:CDEO}
\hat{I}_k[\bm{x}_{k+1}(t'),t] = \bm{M}(t,t_k)\, \bm{x}_{k+1}(t_{k+1}) + \int_{t_k}^{t} dt'\, \bm{M}(t,t')\,\bm{B}(t')\,\bm{x}_{k+1}(t'), \quad t \in [t_k,t_{k+1}].
\end{equation}
\end{widetext}
Since the operator $\hat{I}_k$ is equivalent to the solution operator for DDEs with constant delay in terms of the \emph{Hale-Krasovkii notation} where the time-domain of the state is shifted to the time-domain of the initial function and the position in the physical time is represented by labeling the state intervals \cite{1993hale}, we call it \emph{constant delay evolution operator}.
For convenience, we introduce the following short notation
\begin{equation}
\hat{L} \,\cdot\, = \hat{L}[ \,\cdot\, ,t],
\end{equation}
where $\hat{L}$ stands for any of the above and below defined operators.

\begin{figure}[htp]
\raggedright
\hspace{5mm} a) \hspace{0.18\textwidth} b)\\
\centering
\includegraphics[width=0.21\textwidth]{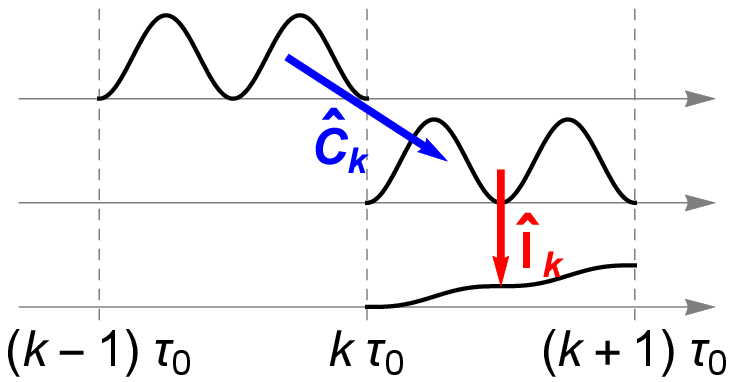}
\includegraphics[width=0.21\textwidth]{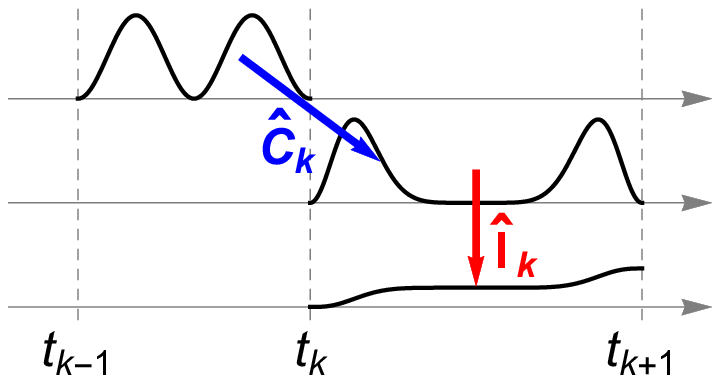}
\colorcaption{\label{fig:mos} Illustration of the decomposition of the solution operator into the sequential action of the access operator $\hat{C}_k$ and the constant delay evolution operator $\hat{I}_k$ for a) constant delay and b) time-varying delay. For constant delay $\tau_0$ the access operator $\hat{C}_k$ reduces to a time-shift by $\tau_0$ and for time-varying delay the access operator changes the length of the state-interval and deforms the state on the time-domain.}
\end{figure}

Finally the evolution of a solution of our linear DDE can be expressed in terms of time-intervals with finite length, where the connection between two states on sequential time-intervals is given by
\begin{equation}
\label{eqn:solop}
\bm{x}_{k+1}(t) = \hat{I}_k\,\hat{C}_k\,\bm{x}_{k}(t')
\end{equation}
and we have derived a method that separates the dynamics of the delay system with time-varying delay into the access map's dynamics represented by the access operators $\hat{C}_k$ and the dynamics of a system with constant delay represented by the constant delay evolution operators $\hat{I}_k$. The method is illustrated in Fig.~\ref{fig:mos}.

\section{\label{sec:finsec} Decomposing the dynamics via Finite section method}

In the following we introduce a method to determine the relaxation rate of the evolution of small volumes, called \emph{average divergence}, under the dynamics of the DDE~\eqref{eqn:sys1} on a finite dimensional subspace of its phase space. The method, which is a generalization of Farmers approach for systems with constant delay \cite{1982farmer} to systems with time-varying delay, is based on the method for calculating Lyapunov exponents given in \cite{2014breda} and uses the presented separation of the solution operator in Sec.~\ref{sec:prelim}.

Let us define the sets $\{\phi_{k i}(t)\}_{i\in\mathbb{N}}$ to be the bases of each canonical direction of the function spaces $\mathcal{F}\left([t_{k-1},t_{k}],\mathbb{R}^d\right)$ and $\{\psi_{k i}(t)\}_{i\in\mathbb{N}}$ to be the corresponding dual sets in the dual spaces $\mathcal{F}^*$ \footnote{Note that in the literature, particularly in \cite{2014breda,1978bernier}, the dynamics of DDEs is represented by a family of solution operators rather on $\mathbb{R}^d\times \mathcal{F}$ than $\mathcal{F}$ because of some mathematical issues defining the solution operator on Lebesgue spaces, for example. But as long as we only consider continuous initial functions we avoid these issues.}. In other words, let the mentioned sets define biorthogonal systems of our spaces $\mathcal{F}$ endowed with a suitable inner product of vectors of the same and of vectors of different spaces $\mathcal{F}$. Then the state $\bm{x}_k(t)$ can be expanded into a series in terms of the basis vectors of the underlying space, i.e.
\begin{equation}
\bm{x}_k(t)=\sum_{i=0}^\infty \bm{q}_{k,i}\, \phi_{k i}(t).
\end{equation}
Thus, $\bm{x}_k(t)$ is represented as an infinite-dimensional vector $\bm{q}_k = \mathrm{cols}(\bm{q}_{k,0},\bm{q}_{k,1},\dots)$ composed of the $d$-dimensional coefficient vectors $\bm{q}_{k,i}$. The $i$-th vector $\bm{q}_{k,i}$ is related to the $i$-th basis function $\phi_{k i}(t)$ and its components are related to the canonical directions of the state $\bm{x}_k(t)$. Furthermore, we define the matrix representations $\bm{I}_k$ and $\bm{C}_k$ of the operators $\hat{I}_k$ and $\hat{C}_k$, respectively, such that the evolution of the state by eq.~\eqref{eqn:solop} is represented by
\begin{equation}
\label{eqn:solopinfmat}
\bm{q}_{k+1} = \bm{I}_k\,\bm{C}_k\,\bm{q}_{k}.
\end{equation}
Obviously, the infinite-dimensional matrices  $\bm{I}_k$ and $\bm{C}_k$ are composed of $(d\times d)$-matrices, whereby each is related to one of the basis functions $\phi_{k i}(t)$. For the following analysis it is convenient to choose matrix representations, i.e., to choose the bases and the inner product, in a way such that shift operators are represented by identity matrices. This is a quite natural assumption, because in our framework the shift operator is implemented by just switching the underlying space $\mathcal{F}\left([t_{k-1},t_{k}],\mathbb{R}^d\right)$. In the case of a constant delay the operators $\hat{C}_k$ become shift operators and the whole dynamics is sufficiently described by the $\hat{I}_k$. One possible definition of such a matrix representation can be found in Appendix~\ref{sec:appmat}.

Let $\bm{I}_k^{(n)}$, $\bm{C}_k^{(n)}$ and $\bm{q}_k^{(n)}$ be the $n$-dimensional approximation of the abovementioned operators and the state at time $t_k$, where $n$ is an integer multiple  of $d$. The $n$-dimensional approximation  $\bm{q}_{N}^{(n)}$ of the state after $N$ time-steps $\bm{x}_{N}(t)$ is now given by
\begin{equation}
\bm{q}_{N}^{(n)} = \prod_{k=0}^{N-1} \bm{I}_k^{(n)} \bm{C}_k^{(n)} \bm{q}_{0}^{(n)}.
\end{equation}

Now let us consider the projection of the dynamics of our linear delay system defined by Eq.~\eqref{eqn:linsys} on $n$-dimensional subspaces of $\mathcal{F}\left([t_{k-1},t_{k}],\mathbb{R}^d\right)$ and analyze the evolution of a small $n$-dimensional initial volume $V^{(n)}_0$ spanned by $n$ linearly independent vectors on such a subspace of the space of initial functions $\mathcal{F}\left([t_{-1},t_{0}],\mathbb{R}^d\right)$. The long term evolution of these vectors and the related volume under the dynamics of Eq.~\eqref{eqn:linsys} in general is dominated by the $n$ largest Lyapunov exponents $\lambda_l$. Thus, the asymptotics of the $n$-dimensional volume $V^{(n)}_{N}$ at time $t_N$ after subsequent projection to the related $n$-dimensional subspace of the underlying space $\mathcal{F}\left([t_{N-1},t_{N}],\mathbb{R}^d\right)$ is given by
\begin{equation}
\label{eqn:volasym}
V^{(n)}_{N} \sim e^{\sum_{l=1}^{n} \lambda_l (t_N-t_0)}.
\end{equation}
Since the long-time evolution of a small volume is in general independent of the specific subspace, we choose as basis vectors for our $n$-dimensional subspace the first $m=n/d$ members of our base $\{\phi_{k i}(t)\}_{i\in\mathbb{N}}$ for each of the $d$ canonical directions of the phase space. So an approximation of the volume evolution follows directly from the finite section approximation of the evolution operator and is given by   
\begin{equation}
\frac{V^{(n)}_{N}}{V^{(n)}_{0}} \approx \left| \det\left( \prod_{k=0}^{N-1} \bm{I}_k^{(n)} \bm{C}_k^{(n)} \right) \right|.
\end{equation}
Utilizing the factorization of the determinant of the product of matrices, we obtain
\begin{equation}
\frac{V^{(n)}_{N}}{V^{(n)}_{0}} \approx \left( \prod_{k=0}^{N-1} \left| \det\left( \bm{I}_k^{(n)} \right) \right| \right) \cdot \left| \det\left( \prod_{k=0}^{N-1} \bm{C}_k^{(n)} \right) \right|.
\end{equation}
The codomain of the access operator $\hat{C}_{k}$ equals the domain of the access operator $\hat{C}_{k+1}$. Hence the matrix product of their finite section approximations can be identified with the finite section approximation of their product. So we consider the product of the access operators $\hat{C}_k$
\begin{equation}
\label{eqn:defopD}
\hat{D}_N = \prod_{k=0}^{N-1} \hat{C}_{k} := \hat{C}_{N-1} \cdots \hat{C}_{1} \hat{C}_0
\end{equation}
and let $\bm{D}_{N}^{(n)}$ be its $n$-dimensional finite section approximation. Furthermore we assume that the convergence of this approximation in terms of $\{\phi_{k i}(t)\}_{i\in\mathbb{N}}$ is well-behaved in the sense that for some unbounded sequence of $N$
\begin{equation}
\det(\bm{D}_{N}^{(n)}) \sim \det\left( \prod_{k=0}^{N-1} \bm{C}_k^{(n)} \right).
\end{equation}
Finally, the volume evolution can be expressed by 
\begin{equation}
\frac{V^{(n)}_{N}}{V^{(n)}_{0}} \approx \left( \prod_{k=0}^{N-1} \left| \det\left( \bm{I}_k^{(n)} \right) \right| \right) \cdot |\det(\bm{D}_{N}^{(n)})|.
\end{equation}
Obviously the volume evolution factorizes into two parts where the first part describes the influence of the successive integrations and the second part describes the influence of the underlying dynamics of the successive access to the initial state $\bm{x}_0(t)$ by the retarded argument $R(t)$.

More expressive than the explicit volume evolution is the related exponential rate, the average divergence $\delta_n$, which can be computed by
\begin{equation}
\label{eqn:sepdiv}
\delta_n = \lim\limits_{N\to\infty} \frac{1}{t_N-t_0} \log\left( \frac{V^{(n)}_{N}}{V^{(n)}_{0}} \right) = \delta_{\bm{I}^{(n)}} + \delta_{\bm{C}^{(n)}}
\end{equation}
and can be separated into the parts
\begin{equation}
\label{eqn:muIn}
\delta_{\bm{I}^{(n)}}= \lim\limits_{N\to\infty} \frac{1}{t_N-t_0} \sum_{k=0}^{N-1} \log \left| \det\left( \bm{I}_k^{(n)} \right) \right|.
\end{equation}
and
\begin{equation}
\label{eqn:muCn}
\delta_{\bm{C}^{(n)}}= \lim\limits_{N\to\infty} \frac{1}{t_N-t_0}  \log \left|\det\left(\bm{D}_{N}^{(n)}\right)\right|.
\end{equation}
The exponential rate $\delta_{\bm{I}^{(n)}}$ describes the volume evolution caused by the successive integrations and the rate $\delta_{\bm{C}^{(n)}}$ describes the volume evolution caused by the successive access to the initial state by the retarded argument $R(t)$. Due to Eq.~\eqref{eqn:volasym}, $\delta_n$ describes the asymptotics of the sum of the $n$ largest Lyapunov exponents. Hence, the asymptotic scaling behavior of the Lyapunov spectrum can be computed by the limit of the difference of the approximations of $\delta_n$ and $\delta_{n-1}$ for large $n$. Just as the average divergence the asymptotic scaling of the Lyapunov spectrum splits into two parts
\begin{equation}
\label{eqn:seplya}
\lambda_n \sim \delta_n - \delta_{n-1} = \lambda_{\bm{I}^{(n)}} + \lambda_{\bm{C}^{(n)}}.
\end{equation}
In the method described above $n$ has to be a rational multiple of the dimension $d$. Thus, Eq.~\eqref{eqn:seplya} can only applied directly, if $d=1$ or if we have found expressions describing the asymptotics of Eq.~\eqref{eqn:muIn} and Eq.~\eqref{eqn:muCn} for large arbitrary $n$. Otherwise, the asymptotic scaling of the Lyapunov spectrum can be approximated by
\begin{equation}
\label{eqn:dseplya}
\lambda_n \sim \frac{1}{d} (\delta_n - \delta_{n-d}).
\end{equation}

Now let us investigate the part $\delta_{\bm{I}^{(n)}}$ of the average divergence, respectively, the influence of the constant delay evolution operators $\hat{I}_k$. As already mentioned the operators $\hat{I}_k$ can be interpreted as evolution operators of one time-step of the linear delay Eq.~\eqref{eqn:linsys} where the variable delay $\tau(t)$ is substituted by the constant delay $\tau_k=t_{k+1}-t_k$. The divergence of their finite section approximation by matrices computed by the Euler method can be determined by Farmer's method \cite{1982farmer}. For simplicity we assume that $\log|\det(\bm{B}(t))|$ is integrable for all $t>t_0$. Thus, we obtain the asymptotic approximation for large $n$
\begin{equation}
\log |\det( \bm{I}_k^{(n)} )| \approx - n \log(n) + O(n).
\end{equation}
With Eq.~\eqref{eqn:muIn} we obtain for the asymptotic scaling of the part $\delta_{\bm{I}^{(n)}}$ of the average divergence 
\begin{equation}
\delta_{\bm{I}^{(n)}} \approx  - \frac{1}{\bar{\tau}} n \log(n) + O(n),
\end{equation}
where $\bar{\tau}$ denotes the arithmetic average, given by
\begin{equation}
\bar{\tau} = \lim\limits_{N\to\infty} \frac{1}{N} \sum_{k=0}^{N-1} \tau_k .
\end{equation}
As a consequence, the asymptotic scaling behavior of the part $\lambda_{\bm{I}^{(n)}}$ of the Lyapunov spectrum, related to the constant delay evolution operators, results in
\begin{equation}
\label{eqn:asymlyaint}
\lambda_{\bm{I}^{(n)}} \sim - \frac{1}{\bar{\tau}} \log(n)
\end{equation}

In this section we have demonstrated that the average divergence and the asymptotic scaling behavior of the Lyapunov spectrum consist of two parts, which are defined by Eq.~\eqref{eqn:sepdiv} and Eq.~\eqref{eqn:seplya}, respectively. One part describes the dynamics related to the constant delay evolution operators and characterizes a fundamental property of every time-delay system. The second part describes the characteristic dynamics of the retarded access, i.e., the deformation of the state-space in the time domain. In Sec.~\ref{sec:specdelayclasses} the above presented method is applied to derive the influence of the dynamics of the retarded access on the asymptotic scaling of the Lyapunov spectrum.


\section{\label{sec:specdelayclasses} Influence of the delay-type on the Lyapunov spectrum}

In this section we relate the delay types identified in Sec.~\ref{sec:class} to the asymptotic scaling behavior of the average divergence and the asymptotic scaling behavior of the Lyapunov spectrum.

\subsection{Conservative delays}
Let us begin with a special case of a conservative delay and define the retarded argument by $R(t)=t-\tau$ with the constant delay $\tau$. With Eq.~\eqref{eqn:compo2} in Sec.~\ref{sec:prelim} we see that the access operators $\hat{C}_k$ reduce to the forward shift operator, shifting the state by the delay $\tau$. Since, the shift is implemented by the change of the underlying space and is represented by an identity matrix as defined in Sec.~\ref{sec:finsec}, the finite section approximation equals the $n$-dimensional identity matrix $\bm{E}_{n}$ in any base.
\begin{equation}
\quad \bm{C}_k^{(n)} = \bm{E}_{n}
\end{equation}
Since the determinant of $\bm{E}_n$ is equal to one we obtain $\det(\bm{D}_{N}^{(n)})=1$ for all $n$. Obviously the part $\delta_{\bm{C}^{(n)}}$ of the average divergence equals zero and according to Eq.~\eqref{eqn:asymlyaint} the asymptotic scaling of the Lyapunov spectrum results in 
\begin{equation}
\label{eqn:lyasclacons}
\lambda_n \sim - C \log(n),
\end{equation}
which is equivalent to the result obtained by Farmer \cite{1982farmer}.

For general conservative delay, let the access map $R(t)$ fulfill Eq.~\eqref{eqn:crit} in Sec.~\ref{sec:class} with irrational rotation number $c=\lim_{j\to\infty} \frac{p_j}{q_j}$. The case of a conservative delay with rational rotation number is automatically included in the following analysis by setting $\frac{p_j}{q_j}=\frac{p}{q}$ and leads to the same result. With Eq.~\eqref{eqn:crit} and Eq.~\eqref{eqn:koopmanopdef} we obtain
\begin{equation}
\label{eqn:acopittir}
\hat{C}^{q_j}\,\bm{x}(t'+p_j) \overset{j\to\infty}{\longrightarrow} \bm{x}(t).
\end{equation}
Hence the $q_j$-th iteration of the Koopman operator $\hat{C}$ converges to the temporal shift, which does not change the shape of the state and the length of the state interval. So it is convenient to calculate the average divergence after $q_j$ time steps and the part $\delta_{\bm{C}^{(n)}}$ can be computed by utilizing the finite section approximation of the operator $\hat{D}_{q_j}$ defined by Eq.~\eqref{eqn:defopD}. With Eq.~\eqref{eqn:acopittir} it is clear that the operator $\hat{D}_{q_j}$ converges to the temporal shift by $p_j$ and since the shift is represented by the identity matrix, the finite section approximation of $\hat{D}_{q_j}$ converges to the identity matrix
\begin{equation}
\lim\limits_{j\to\infty} \bm{D}_{q_j}^{(n)} = \bm{E}_{n}.
\end{equation}
Obviously the determinant $\det(\bm{D}_{q_j}^{(n)})$ related to the volume evolution caused by the access operator tends to one if we send $j\to\infty$. The corresponding parts $\delta_{\bm{C}^{(n)}}$ and $\lambda_{\bm{C}^{(n)}}$ of the average divergence and the Lyapunov spectrum, respectively, vanish and the Lyapunov spectrum shows the logarithmic scaling behavior characteristic to systems with constant delay
\begin{equation}
\lambda_n \sim - C \log(n),
\end{equation}
which is equivalent to Eq.~\eqref{eqn:lyasclacons} as one expects due to the invariance of the Lyapunov exponents under suitable timescale transformations.

\subsection{Dissipative delays}

Let us assume $R(t)$ to be the lift of a circle map with rotation number $\frac{p}{q}$ that fails the criterion Eq.~\eqref{eqn:crit}. Poincare's classification for circle maps implies the existence of points $t^*$ with $R^q(t^*)+p=t^*$, in other words $q$-periodic orbits \cite{1997katok}. For convenience and without any loss of generality we set the initial time $t_0$ to an arbitrary periodic point $t^*$, leading to a periodic sequence of interval lengths for the method of steps. The periodicity of the delay $\tau(t)$ and the definition of the operators $\hat{C}_{k}$ imply that they are in some sense periodic with respect to $k$. The operator $\hat{C}_{k+q}$ is almost identical to the operator $\hat{C}_{k}$ with the difference that the domain of the elements of the underlying space is shifted by $p$ relative to the domain of the elements of the spaces where $\hat{C}_{k}$ lives. Hence $\hat{C}_{k+q}$ is connected to $\hat{C}_{k}$ by
\begin{equation}
\hat{C}_{k+q}=\hat{S}^{-1}_p\, \hat{C}_{k}\,\hat{S}_p,
\end{equation}
where the temporal displacement is represented by the shift operator
\begin{equation}
\label{eqn:shiftdef}
\hat{S}_p\, \bm{x}_k(t') = \bm{x}_k(t+p).
\end{equation}
Since the length of the state intervals repeats after $q$ time-steps, it will be convenient to compute the average divergence after $j\,q$ time-steps. Therefore we compute the operator $\hat{D}_{jq}$ using our definitions above and obtain
\begin{equation}
\hat{D}_{jq} = \prod_{l=0}^{j-1} \hat{S}^{-1}_{lp} \left(\prod_{k=0}^{q-1} \hat{C}_{k}\right) \hat{S}_{lp} = \prod_{l=0}^{j-1} \hat{S}^{-1}_{lp} \hat{D}_q \hat{S}_{lp},
\end{equation}
where $\hat{D}_q$ is defined as in Eq.~\eqref{eqn:defopD}.
Due to the factorization of the determinant and the fact that the shift operator is represented by the identity matrix as described for conservative delay, the determinant of the $n$-dimensional finite section approximation of $\hat{D}_{jq}$ simplifies to 
\begin{equation}
\det\left(\bm{D}_{j q}^{(n)}\right) = \det\left(\bm{D}_{q}^{(n)}\right)^j.
\end{equation}
The corresponding operator $\hat{D}_q$ can be interpreted as the restriction of the Koopman operator related to the $q$-th iteration of the access map $R(t)$ to an interval, whose boundaries are given by two succeeding points of a periodic orbit of $R(t)$. Hence the symbol of the operator $\hat{D}_q$ equals $R^q(t)$. Since the state after $q$ iterations of the solution operator is shifted in time by $p$ relative to the initial state, the part $\delta_{\bm{C}^{(n)}}$ of the average divergence has to be rescaled by $p$ and according to Eq.~\eqref{eqn:dseplya} and Eq.~\eqref{eqn:asymlyaint} the asymptotic Lyapunov spectrum is given by
\begin{equation}
\label{eqn:lyadiss}
\lambda_n \sim - C \log(n) + \frac{1}{d\,p}\,\log \left|\frac{\det\left(\bm{D}_{q}^{(n)}\right)}{\det\left(\bm{D}_{q}^{(n-d)}\right)}\right|.
\end{equation}

In the further analysis we show that a dissipative delay leads to a linear asymptotic scaling behavior of the Lyapunov spectrum, which can be made plausible by the following property of the eigenfunctions and eigenvalues of the Koopman operator. So let $\bm{\xi}(t)$ be an eigenfunction of the operator $\hat{D}_q$ with the corresponding eigenvalue $a$. Since $\hat{D}_q$ maps a function to a function space whose member's domain is shifted by $p$, the terms ``eigenvalue'' and ``eigenfunction'' should be interpreted as solutions of
\begin{equation}
a\, \bm{\xi}(t-p) = \hat{D}_q \bm{\xi}(t').
\end{equation}
With the short calculation
\begin{equation}
\hat{D}_q\,(\bm{\xi}(t'))^j=(\bm{\xi}(R^q(t)))^j=(a\,\bm{\xi}(t-p))^j=a^j\,(\bm{\xi}(t-p))^j
\end{equation}
it follows directly that $(\bm{\xi}(t))^n$ is also an eigenfunction of the operator $\hat{D}_q$ but with the corresponding eigenvalue $a^j$, whereby a constant $\bm{\xi}(t)=\bm{\xi}^*$ is always an eigenfunction with corresponding eigenvalue $1$. In other words the eigenfunctions and eigenvalues of a Koopman operator form a semi-group \cite{2012budisic}. We assume that there is only one $0<a<1$ defining the whole spectrum by the semi-group property, which holds if the Koopman operator $\hat{D}_q$ is compact and $a$ equals the slope of the symbol $R^q(t)$ of $\hat{D}_q$ at the sole attractive fixed point $t^*$ inside the state interval \cite{1975caughran}, where ``fixed point'' means in our case a point fulfilling $t^*=R^q(t^*)+p$. Thus, we first consider for simplicity that $p=1$ and that the reduced access map $R(t) \mod 1$ has a unique attractive $q$-periodic orbit. The operator $\hat{D}_q$ acts independently and identically at each of the $d$ components of the initial state $\bm{x}_0(t)$, which are approximated in terms of the first $m$ members of the basis $\{\phi_{k i}(t)\}_{i\in\mathbb{N}}$. Hence, all eigenvalues occur with multiplicity $d$ and with the assumption that the determinant of $\bm{D}_{q}^{(n)}$ is now given by the product of the largest $n=m\,d$ eigenvalues of $\hat{D}_q$ counting multiplicity, we obtain
\begin{equation}
\det(\bm{D}_{q}^{(n)})=\prod_{k=0}^{m-1} a^{d\,k} = a^{\frac{d}{2}m(m-1)} = a^{\frac{1}{2d}n(n-d)}.
\end{equation}
From the definitions of the part $\delta_{\bm{C}^{(n)}}$ of the average divergence and the part $\lambda_{\bm{C}^{(n)}}$ of the Lyapunov spectrum, i.e., the part related to $\hat{D}_q$ in Eq.~\eqref{eqn:lyadiss}, and with $a \equiv e^{\hat{\alpha}}$ follows directly
\begin{equation}
\label{eqn:compKooplya}
\delta_{\bm{C}^{(n)}}=\frac{1}{2d}n(n-d)\,\hat{\alpha}, \quad \lambda_{\bm{C}^{(n)}} \sim \frac{n}{d} \hat{\alpha}.
\end{equation}

\begin{figure}[htp]
\raggedright
\hspace{5mm} a)\\
\centering
\includegraphics[width=0.35\textwidth]{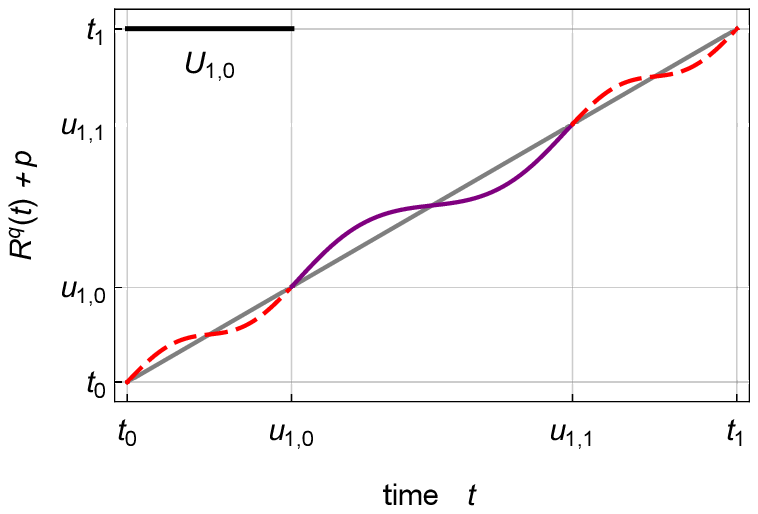}\\
\raggedright
\hspace{5mm} b)\\
\centering
\includegraphics[width=0.35\textwidth]{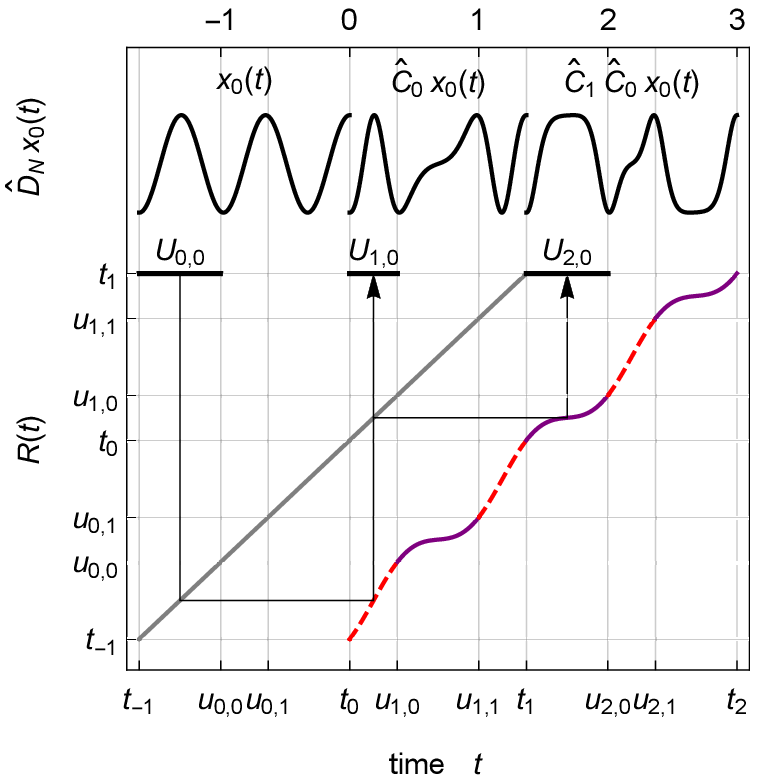}
\colorcaption{\label{fig:qperiosep} a) Separation of the state intervals into the basins of attraction $U_{k,j}=[u_{k,j-1},u_{k,j}]$, where $u_{k,p-1}=t_k$, of the attractive fixed points of the $q$-th iteration $R^q(t)$ of an exemplary access map $R(t)$ with rotation number $\frac{p}{q}=\frac{3}{2}$. b) Evolution of a function $x_0(t)$ inside the $U_{k,j}$ by the successive application of the access operators $\hat{C}_k$. The basic types of symbols of $\hat{C}_{k,j}$, i.e., segments of $R(t)$, are illustrated by the alternating dashed an solid lines.}
\end{figure}

For an access map with $p>1$, where the reduced map has again a unique attractive periodic orbit, the argumentation above has to be slightly modified since $R^q(t)$ has $p$ attractive fixed points inside the state interval. We separate the state interval $[t_{k-1},t_k]$ into the basins of attraction $U_{k,j}=[u_{k,j-1},u_{k,j}]$, where $u_{k,p-1}=t_k$, of the fixed points of $R^q(t)$ as illustrated in Fig.~\ref{fig:qperiosep}. Therefore, we define the restriction $\bm{x}_{k,j}(t)$ of the state $\bm{x}(t)$, the restriction $\hat{C}_{k,j}$ of the Koopman operator $\hat{C}$ and the restriction $\hat{D}_{q,j}$ of the operator $\hat{D}_q$ by 
\begin{subequations}
\begin{align}
\bm{x}_{k,j}(t) &= \bm{x}(t),& \quad t\in [u_{k,j-1},u_{k,j}]\\
\hat{C}_{k,j}\, \bm{x}_{k,j}(t') &= \bm{x}_{k,j}(R(t)),& \quad t\in [u_{k,j-1},u_{k,j}]\\
\hat{D}_{q,j}\, \bm{x}_0(t') &= \bm{x}_0(R^q(t)),& \quad t\in [u_{q,j-1},u_{q,j}].
\end{align}
\end{subequations}
With Fig.~\ref{fig:qperiosep}b it is easy to understand that the Koopman operator $\hat{C}$ maps the basin of attraction $U_{k,j}$ to the basin of attraction $U_{k+1,j}$ and that the restrictions $\hat{C}_{k,j}$ of the Koopman operator are basically given by $q$ different Koopman operators shifted in time, which correspond to the symbols given by $q$ different segments of $R(t)$, repeating with period $q$ in $k$ and $j$. Thus, we can simplify our analysis by introducing $q$ basic operators $\tilde{C}_0,\tilde{C}_1,\dots,\tilde{C}_{q-1}$, which are connected to the operators $\hat{C}_{k,j}$ by simple time shifts, i.e.
\begin{equation}
\hat{C}_{k,j} = \hat{S}^{-1}_{u_{k+1,j}}\, \tilde{C}_{(k\,p+j) \;\mathrm{mod}\; q}\, \hat{S}_{u_{k,j}},
\end{equation}
where the operators $\hat{S}$ are shift operators as defined by Eq.~\eqref{eqn:shiftdef}. In the same way, we represent the operators $\hat{D}_{q,j}$ by
\begin{eqnarray}
\hat{D}_{q,j} &=& \hat{S}^{-1}_{u_{q,j}} \left( \prod_{k=0}^{q-1} \tilde{C}_{(k\,p+j) \;\mathrm{mod}\; q} \right) \hat{S}_{u_{0,j}}\nonumber\\
&=& \hat{S}^{-1}_{u_{q,j}} \tilde{D}_{q,j} \hat{S}_{u_{0,j}}
\end{eqnarray}
Since the shift operators only change the domain of the members of the domain and codomain of the operators, the spectrum of $\hat{C}_{k,j}$ equals the spectrum of the corresponding basic operator and the spectrum of $\hat{D}_{q,j}$ equals the spectrum of $\tilde{D}_{q,j}$. With Fig.~\ref{fig:qperiosep}b it is easy to see that the operators $\tilde{D}_{q,j}$ are given by
\begin{equation}
\tilde{D}_{q,j}=\tilde{C}_{j} \tilde{C}_{j+1} \cdots \tilde{C}_{q-1} \tilde{C}_{0} \tilde{C}_{1} \cdots \tilde{C}_{j-1},
\end{equation}
such that $\tilde{D}_{q,j+1}$ can be obtained from $\tilde{D}_{q,j}$ by switching the order of the operator product between $\tilde{C}_{j}$ and the remaining product of operators. With the fact given in \cite{barnes_common_1998} that switching the order of the product of two operators does not change the spectrum of the total operator, except the eigenvalues equal to zero, together with the fact that the spectrum of $\hat{D}_{q,j}$ equals the spectrum of $\tilde{D}_{q,j}$, we obtain
\begin{equation}
\label{eqn:Dspec}
\mathrm{spec}(\hat{D}_{q,j}) \backslash \{ 0 \} = \mathrm{spec}(\hat{D}_{q,j'}) \backslash \{ 0 \},\quad \forall\; j,j'.
\end{equation}
Since the Koopman operator maps $U_{k,j}$ to $U_{k+1,j}$ and each state interval consists of $p$ basins of attraction, the operator $\hat{D_q}$ can be represented by the operator valued matrix $\mathrm{diag}(\hat{D}_{q,0},\hat{D}_{q,1},\dots,\hat{D}_{q,p-1})$. Hence and with Eq.~\eqref{eqn:Dspec} it is clear that the spectrum of $\hat{D}_q$ equals the spectrum of all of the $\hat{D}_{q,j}$ and each eigenvalue of $\hat{D}_{q,j}$ is an eigenvalue of $\hat{D}_{q}$ with multiplicity $p$. At this point, we expand the restrictions $\bm{x}_{k,j}(t)$ of the state in terms of some basis functions as done in Sec.~\ref{sec:finsec}, with the difference that the domain of the basis functions is now given by the basin of attraction $U_{k,j}$. By doing this, we obtain coefficient vectors $\bm{q}_{k,j,i}$ corresponding to the $i$-th basis function. The state $\bm{x}_k(t)$ inside the $k$-th state interval now can be represented by an infinite-dimensional vector composed of the $d p$-dimensional coefficient vectors $\bm{q}_{k,i}=\mathrm{cols}(\bm{q}_{k,0,i},\bm{q}_{k,1,i},\dots,\bm{q}_{k,p-1,i})$, which correspond to the $i$-th basis function of each basin of attraction $U_{k,j}$ inside the $k$-th state interval. Due to the block-diagonal structure of $\hat{D}_q$ the matrix representation of the operator $\hat{D}_q$ with respect to the mentioned expansion of the state is given by a block-diagonal matrix, where the $p$ blocks are the matrix representations of the operator and $\hat{D}_{q,j}$, respectively. Consequently, the related $n=m d p$-dimensional finite section approximation $\bm{D}_{q}^{(n)}$, with respect to the first $m$ basis functions of each of the $p$ basins of attraction, takes the form 
\begin{eqnarray}
\bm{D}_{q}^{(n)}&=&\mathrm{diag}\left(\bm{D}_{q,0}^{(md)},\bm{D}_{q,1}^{(md)},\dots,\bm{D}_{q,p-1}^{(md)}\right),
\end{eqnarray}
where the $\bm{D}_{q,j}^{(md)}$ are the $md$-dimensional finite section approximations of $\hat{D}_{q,j}$. With the assumption that the operators $\hat{D}_{q,j}$ are compact, together with Eq.~\eqref{eqn:Dspec} and the results for the case $p=1$, it follows that $\mathrm{spec}(\hat{D}_{q})=\mathrm{spec}(\hat{D}_{q,j})=\{1,a,a^2,\dots\}\cup \{0\}$, where $a$ denotes the unique slope of $R^q(t)$ at the attractive fixed points and all eigenvalues of $\hat{D}_q$ occur with multiplicity $d p$. With the assumption that the determinant of $\bm{D}_{q}^{(n)}$ is now given by the product of the largest $n=m d p$ eigenvalues of $\hat{D}_q$ counting multiplicity, we obtain
\begin{equation}
\det(\bm{D}_{q}^{(n)})=\prod_{k=0}^{m-1} a^{d p \,k} = a^{\frac{d p}{2}m(m-1)} = a^{\frac{1}{2 d p}n(n-d p)}
\end{equation}
Finally, with $a\equiv e^{\hat{\alpha}}$ we obtain for the part $\delta_{\bm{C}^{(n)}}$ of the average divergence and for the part $\lambda_{\bm{C}^{(n)}}$ of the asymptotic scaling of the Lyapunov spectrum
\begin{equation}
\label{eqn:compKooplyagenp}
\delta_{\bm{C}^{(n)}}=\frac{1}{2 d p^2}n(n-d p)\,\hat{\alpha}, \quad \lambda_{\bm{C}^{(n)}} \sim \frac{n}{d p^2} \hat{\alpha},
\end{equation}
respectively. Hence, the asymptotically linear scaling behavior of the Lyapunov spectrum is directly connected to the semi-group property of the Koopman eigenfunctions. Note that for general access maps $R(t)$ possessing multiple attractive $q$-periodic orbits Eq.~\eqref{eqn:Dspec} does not apply and there is no unique slope $a$ of the attractive fixed points of $R^q(t)$. So there are many ways to combine the $a_l$ corresponding to the fixed points $t^*_l$ by multiplying powers of the related eigenfunctions $\bm{\xi}_l(t)$ using their semi-group property. Thus, the part $\lambda_{\bm{C}^{(n)}}$ of the Lyapunov spectrum and its asymptotic slope $\hat{\alpha}$ should depend on all of the $a_l$. In general the operator $\hat{D}_{q}$ is also not compact. Even the exemplary access map in Fig.~\ref{fig:qperiosep} leads to a non-compact Koopman operator due to the repulsive fixed points of $R^q(t)$ at the boundaries of $U_{k,j}$. However, the semi-group property of the eigenfunctions and eigenvalues of the Koopman operator is valid for all maps and the method described above was successfully applied at DDEs with sawtooth-shaped delay \cite{2016otto}, which leads to a compact operator $\hat{D}_{q}$ in the space of analytic functions and can be seen as the limiting case where the slope at the repulsive fixed points of $R^q(t)$ is infinite. A further discussion about this problem is made at the end of this section.

In the following we quantitatively analyze the average divergence and the asymptotic scaling behavior of the Lyapunov spectrum for delay systems with dissipative delay. Since the contribution of the constant delay evolution operator was already investigated in Sec.~\ref{sec:finsec} we concentrate on the contribution by the access operator or Koopman operator, respectively. So it is sufficient to consider the simplest delay system given by
\begin{equation}
\label{eqn:lameDDE}
\dot{x}(t) = x(R(t)).
\end{equation}

\begin{figure}[htp]
\begin{minipage}{0.26\textwidth}
\raggedright
\hspace{5mm} a)\\
\centering
\includegraphics[width=\textwidth]{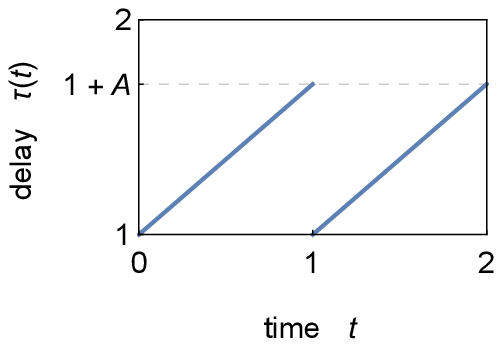}
\end{minipage}
\begin{minipage}{0.16\textwidth}
\raggedright
\hspace{5mm} b)\\
\centering
\includegraphics[width=\textwidth]{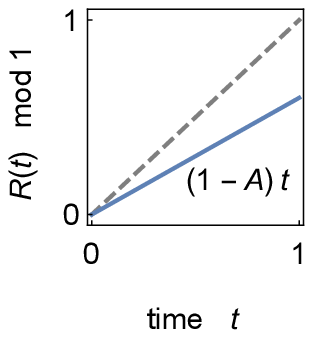}
\end{minipage}
\colorcaption{\label{fig:sawdelay} a) Sawtooth-shaped delay and b) corresponding reduced access map (solid) and bisectrix (dashed)}
\end{figure}

Firstly we analyze a simple exemplary system with dissipative delay where the asymptotic scaling behavior of the Lyapunov spectrum can be computed analytically and which is a special case of the systems with sawtooth-shaped delay analyzed in the supplemental material of \cite{2016otto}. We consider the DDE~\eqref{eqn:lameDDE} with a sawtooth-shaped delay. The definitions of the delay $\tau(t)$ and the retarded argument $R(t)$ read
\begin{equation}
\tau(t)= 1 + A (t-\lfloor t \rfloor) \quad \dot{R}(t) = 1-A \text{, if } t \notin \mathbb{N}.
\end{equation}
With Fig.~\ref{fig:sawdelay}b it is easy to see that for $0 < A < 1$ the reduced access map corresponding to the retarded argument is a simple linear map possessing an attractive fixed point at the origin. If we set the initial time $t_0=0$, the boundaries $t_k$ of our state intervals related to the method of steps and given by the inverse iteration of the access map are equal to the natural numbers and the interval length is equal to the delay's period. So the access operators $\hat{C}_k$ are given by
\begin{equation}
\hat{C}_k x_k(t') = x_k((1-A)(t-k)+k-1), \quad t \in [k,k+1]
\end{equation}
and the constant delay evolution operators $\hat{I}_k$ are given by
\begin{equation}
\hat{I}_k x_{k+1}(t') = x_{k+1}(k+(1-A)^{-1}) + \int_{k}^t x_{k+1}(t')\, dt'.
\end{equation}
Since the length of the state intervals is equal to the period of the delay, the solution operator $\hat{I}_k\hat{C}_k$ is a monodromy operator known from \emph{Floquet theory} and all information about the dynamics can be obtained by analyzing one time-step. In the following we derive a matrix representation of the monodromy operator utilizing the method introduced in Sec.~\ref{sec:prelim} and \ref{sec:finsec} together with the definition of the matrix representation in Appendix~\ref{sec:appmat}. A convenient choice of the biorthogonal bases needed for the finite section approximation are the monomial base and the corresponding adjoint base consisting of the $i$-th derivatives $\delta^{(i)}(t)$ of the delta distribution $\delta(t)$, i.e., we expand the state $x_k(t)$ into a Taylor series around $t=k$. So we set
\begin{subequations}
\begin{align}
\phi_{k i}(t) &= (t-k+1)^i,\\
\psi_{k i}(t) &= \frac{(-1)^i}{i!} \delta^{(i)}(t-k+1) \text{ and}\\
w(t) &= 1.
\end{align}
\end{subequations}
The $n$-dimensional finite section approximation with respect to the monomial base of the constant delay evolution operators $\hat{I}_k$ is computed by Eq.~\eqref{eqn:CDEOmat} in Appendix~\ref{sec:appmat} and results in
\begin{equation}
\bm{I}_{k}^{(n)} =
\begin{pmatrix}
1 & \frac{1}{(1-A)} & \frac{1}{(1-A)^2} & \dots & \frac{1}{(1-A)^{n-1}}\\
1 & 0 & 0 & \dots & 0\\
0 & \frac{1}{2} & 0 & \dots & 0\\
\vdots & \ddots & \ddots & \ddots & \vdots \\
0 & \dots & 0 & \frac{1}{n-1} & 0
\end{pmatrix},
\end{equation}
where the first row represents the summation of the value of the state at the interval endpoint $x_k(k)$, ensuring the continuity of the solution $x(t)$. The lower rows are the matrix representation of the integration in the monomial base. In our simple example the $n$-dimensional finite section approximation of the access operators $\hat{C}_k$ computed utilizing Eq.~\eqref{eqn:cproj} in Appendix~\ref{sec:appmat} reduces to a diagonal matrix and is given by
\begin{equation}
\bm{C}_k^{(n)} = \mathrm{diag}(1,1-A,(1-A)^2,\dots,(1-A)^{n-1}).
\end{equation}
Due to the periodicity of the system and the solution operator, $\bm{I}_k^{(n)}$ and $\bm{C}_k^{(n)}$ are time-independent and the two parts of the average divergence $\delta_n$ reduce to the logarithm of the determinant of the abovementioned operators. The part $\delta_{\bm{I}^{(n)}}$ of the average divergence related to the constant delay evolution operator is now given by
\begin{equation}
\delta_{\bm{I}^{(n)}} = \log\left|\det{\left(\bm{I}_{k}^{(n)}\right)}\right| = -\log((n-1)!) - (n-1) \log (1-A).
\end{equation}
Utilizing Stirling's approximation of the factorial leads to
\begin{equation}
\delta_{\bm{I}^{(n)}} \approx -(n-1) \log (n-1) + n - 1 - (n-1) \log (1-A).
\end{equation}
The part $\delta_{\bm{C}^{(n)}}$ of the average divergence, which represents the contribution by the access operator or the Koopman operator, respectively, results in
\begin{equation}
\delta_{\bm{C}^{(n)}} = \log\left|\det{\left(\bm{C}_{k}^{(n)}\right)}\right| = \frac{1}{2} n(n-1) \log (1-A),
\end{equation}
which is equivalent to Eq.~\eqref{eqn:compKooplya}. Thus we obtain for the asymptotic scaling of the Lyapunov spectrum
\begin{equation}
\label{eqn:lyaspecsawtooth}
\lambda_n \sim -\log n + n\log (1-A).
\end{equation}

So we have shown by a straightforward calculation that the Lyapunov spectrum of our simple delay system with dissipative sawtooth-shaped delay is asymptotically linear in the limit of large Lyapunov indices. Furthermore it has to be noticed that in this special case the asymptotic slope of the Lyapunov spectrum equals the logarithm of the slope of the access map at the attractive fixed point, i.e., it equals the access map's Lyapunov exponent.

\begin{figure}[htp]
\begin{minipage}{0.26\textwidth}
\raggedright
\hspace{5mm} a)\\
\centering
\includegraphics[width=\textwidth]{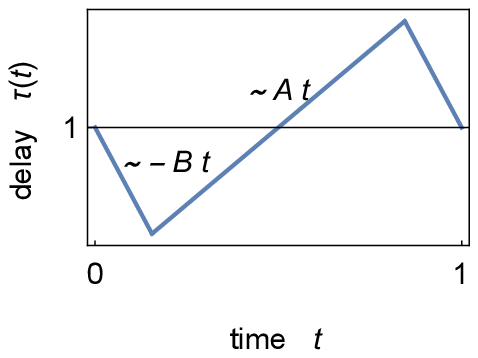}
\end{minipage}
\begin{minipage}{0.18\textwidth}
\raggedright
\hspace{5mm} b)\\
\centering
\includegraphics[width=\textwidth]{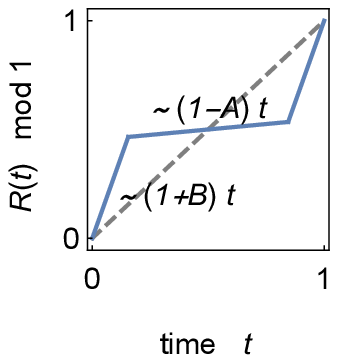}
\end{minipage}
\colorcaption{\label{fig:tridelay} a) Triangle delay characterized by two slopes $A$ and $B$ with b) corresponding reduced access map (solid) and bisectrix (dashed)}
\end{figure}

In the following we analyze the same simple system Eq.~\eqref{eqn:lameDDE} with a more general delay. It is a known fact that sufficiently smooth iterated maps with the same structure of hyperbolic $q$-periodic orbits are topological conjugate, where the slope of the $q$-th iteration $R^q(t)$ of the map $R(t)$ at the periodic points is invariant under differentiable conjugacies \cite{1967smale}. This leads to the conjecture that the asymptotic slope of the Lyapunov spectrum is only influenced by the slopes of $R^q(t)$ at the periodic points of the access map. Hence, we choose a triangular shaped delay as shown in Fig.~\ref{fig:tridelay}a, where the slopes at the attractive and repulsive fixed points can be specified by the parameters $A$ and $B$, respectively. For the further analysis the mean delay is set to one. Thus the reduced access map has an attractive fixed point at $t=1/2$ and repulsive fixed points at the interval boundaries, which are identified due to the modulo operation.

\begin{figure}[htp]
\raggedright
\hspace{5mm} a)\\
\centering
\includegraphics[width=0.4\textwidth]{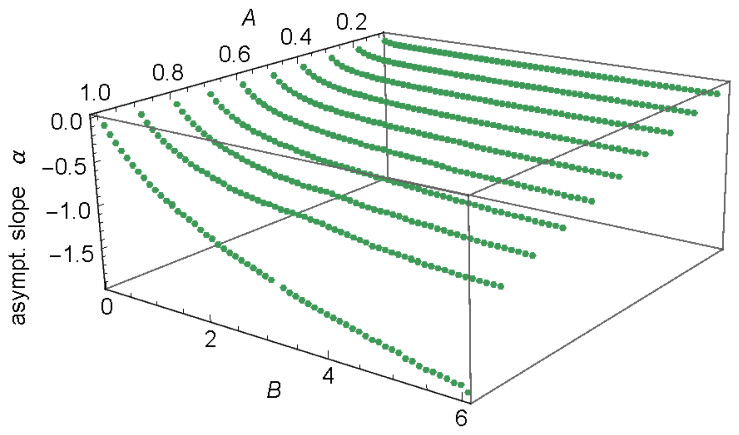}\\
\raggedright
\hspace{5mm} b)\\
\centering
\includegraphics[width=0.4\textwidth]{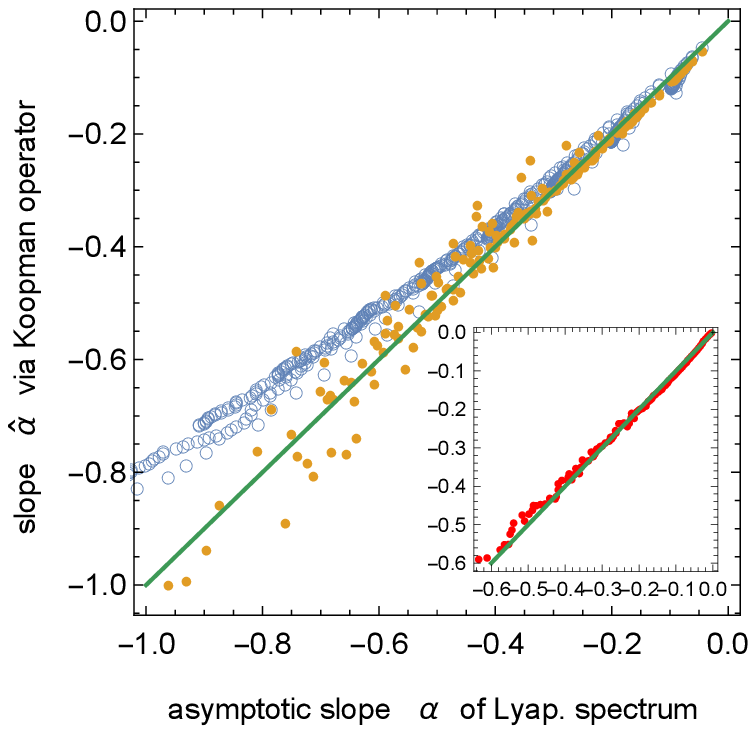}
\colorcaption{\label{fig:tridelayslope} a) Asymptotic slope of the Lyapunov spectrum vs. delay parameters $A$ and $B$ and b) comparison of the asymptotic slope from the finite section approximation of the Koopman operator using an orthonormal base (empty circles) and the Lyapunov base (dots) vs. direct measurement at the Lyapunov spectrum (line), for all $A$ and $B$ from a). The inset shows the same analysis, whereby the Lyapunov base was used, for sine-shaped delay $\tau(t)=1+\frac{A_s}{2\pi}\sin(2\pi\,t)$ under variation of $A_s$.}
\end{figure}

The Koopman operator corresponding to this access map is not compact due to the presence of both attractive and repulsive fixed points. Thus the results above cannot be applied directly and therefore we perform the following numerical analysis. Firstly we have computed the Lyapunov spectra via the semi-discretization method for different values of the parameters $A$ and $B$ and measured the asymptotic slope $\alpha$ of the Lyapunov spectra. The measured slopes are illustrated in Fig.~\ref{fig:tridelayslope}a. It is easy to see that the asymptotic slope of the Lyapunov spectrum depends on both parameters of the delay $A$ and $B$, which justifies our statement that it depends on both slopes of the fixed points of the symbol $R^q(t)$ of the operator $\hat{D}_q$. Additionally we computed the asymptotic slope $\hat{\alpha}$ of the part $\lambda_{\bm{C}^{(n)}}$ of the asymptotic Lyapunov spectrum using the approach presented in Sec.~\ref{sec:finsec}, together with the definition of the matrix representations of the involved operators given in Appendix~\ref{sec:appmat}. For the finite section approximation of the corresponding Koopman operator we choose the Fourier cosine base $\phi_i(t)\in \{1,\sqrt{2}\cos(\pi\,t),\sqrt{2}\cos(2\pi\,t),\dots\}$ on the one hand as a representative example of an orthonormal base, and on the other hand we choose the left eigenfunctions of the monodromy operator corresponding to Eq.~\eqref{eqn:lameDDE} for $\psi_{ki}(t)$ and the corresponding right eigenfunctions or the Lyapunov vectors, respectively, for $\phi_{ki}(t)$, which gives us an appropriate biorthogonal system, which we call \emph{Lyapunov base}. In Fig.~\ref{fig:tridelayslope}b the asymptotic slopes $\hat{\alpha}$ computed by the finite section approach are plotted versus the related slopes $\alpha$ resulting from the direct measurement at the Lyapunov spectrum. So the values resulting from the finite section approximation of the Koopman operator are correct if the points are located at the bisectrix. The approach leads to the correct results if the finite section approximation of the Koopman operator is made with respect to the Lyapunov base corresponding to the specific delay and access map. However, the finite section approach fails for orthonormal bases similar to the Fourier base due to specific structural properties of the Lyapunov vectors, which are discussed in Sec.~\ref{sec:vec}.

Note that the abovementioned results are not restricted to access maps that possess only fixed points. The method is constructed to relate the Koopman operator corresponding to the $q$-th iteration of the access map, where $q$ denotes the period of the periodic orbits of the access map, to the part $\delta_{\bm{C}^{(n)}}$ of the average divergence and the part $\lambda_{\bm{C}^{(n)}}$ of the asymptotic Lyapunov spectrum. Consequently, all dissipative delays that are related to an invertible retarded argument lead to a linear asymptotic scaling behavior of the Lyapunov spectrum. Furthermore the method is not restricted to simple systems like Eq.~\eqref{eqn:lameDDE}. The specific dynamics of the system only influences the properties of the constant delay evolution operator $\hat{I}_k$, whereby the access operator $\hat{C}_k$ and the corresponding Koopman operator only depends on the delay. In the case of a system admitting more complicated, especially chaotic, dynamics one has to take into account that the Lyapunov base changes for each state interval, since the direction of the Lyapunov vectors changes for each state interval. Hence, the computation of the finite section approximation of the Koopman operator gets slightly more complex than for the simple system analyzed above.

\subsection{\label{sec:disc} Discussion}

As we have shown by analyzing the average divergence and the asymptotic scaling behavior of the Lyapunov spectrum, the dynamics of time delay systems is not only influenced by the integral operator but also influenced by the Koopman operator, which describes a dynamical system from the observables point of view \cite{2012budisic}. In the case of delay systems with time-varying delay the dynamical system of interest is given by the access map and the observable is given by the state of the delay system inside a proper interval of the method of steps. Due to the evolution of small volumes on a finite-dimensional subspace of the infinite-dimensional state space the spectral properties of the Koopman operator are connected to the asymptotic scaling behavior of the Lyapunov spectrum, as we have shown before. But there are still open questions on the separate calculation of the spectrum of the Koopman operator.

The spectrum of a Koopman operator depends on the dynamics of the underlying map and can be computed analytically, for example, for maps with a unique (attractive) fixed point, as mentioned before, and for purely expanding circle maps \citep{slipantschuk_analytic_2013,*bandtlow_spectral_2016}. In our case of invertible access maps there are only two possible types of dynamics, which determine our delay types, conservative and dissipative delay. If the access map preserves some measure, the corresponding Koopman operator is unitary with respect to this measure cf. \cite{1931koopman,1989petersen}. Roughly speaking, on the attractor of the map the Koopman operator just mixes up the values of the observable that are assigned to the attractor manifold. For conservative delay we have implicitly shown that the Koopman operator corresponding to the access map is unitary with respect to some non-singular measure, because the limit operator of some special sequence of iterations of the related Koopman operator is the identity. In the case of dissipative delay the attractor only consists of the periodic points or fixed points of the access map. Indeed, the corresponding Koopman operator is unitary with respect to the Dirac measure but the related space is only a finite dimensional subspace of the infinite-dimensional state-space of our DDE, hence less usefull. Except for the spectral analysis of the adjoint of the Koopman operators corresponding to circle maps in \cite{2002hotzel}, there is a lack of literature about this special topic. However, considering the Koopman operator acting on the space of bounded functions $L^\infty$ as the adjoint of the Frobenius-Perron operator acting on $L^1$ is not the right framework for the present problem and leads to wrong results in the case of dissipative delays. There is no need to assume the eigenfunctions of the Koopman operator to be bounded. By doing this one restricts the spectrum of the Koopman operator to the unit circle as shown in \cite{2002hotzel}, even in the case of dissipative delays. The part $\delta_{\bm{C}^{(n)}}$ of the average divergence would be equal to one, which contradicts the analytical and numerical results we derived above.

If a dynamical system posesses hyperbolic fixed points and is sufficiently smooth, the related Koopman operator is well described around this fixed points by the Koopman operator corresponding to the linear approximation of the system, since there is a topological conjugacy between the non-linear system and its linear approximation, which is known as \emph{Hartman-Grobman theorem}, cf. \cite{1960hartman,1959grobman}. The authors of \cite{2013lan} extend the Hartman-Grobman theorem and show that the aforementioned conjugacy exists in the whole basin of attraction if the system is sufficiently smooth and its linear approximation has only eigenvalues with modulus smaller (or all greater) than one. In this case the eigenfunctions of the Koopman operator can be constructed using generalized Laplace averages as shown in \cite{2014mohr} and its spectrum consists of powers of the eigenvalues of the system's linearization due to the semi-group property of the eigenfunctions and eigenvalues of the Koopman operator. These results substantiate our conjecture about the linear scaling behavior of the part $\lambda_{\bm{C}^{(n)}}$ of the Lyapunov spectrum in the case of dissipative delays, which can be proven analytically for our simple linear DDE with a sawtooth-shaped delay that is related to an access map with an attractive fixed point. However, this is not the whole truth since invertible one-dimensional iterated maps possess one repulsive fixed point or periodic orbit per each attractive fixed point or periodic orbit, and the linearization around an attractive fixed point will always be a bad approximation near the corresponding repulsive fixed point. Due to the lack of an analytic approach in this general case, we have done some numerical analysis and are able to justify that the Koopman operator causes the linear scaling behavior of the Lyapunov spectrum. The cause of the deviations of the finite section approach for orthonormal bases is explained in Sec.~\ref{sec:vec}. 

\section{\label{sec:vec} Lyapunov vectors}

For a more detailed analysis of the differences between the two delay classes, we present some qualitative and quantitative properties of the covariant Lyapunov vectors, which can be computed by the method given in \cite{2007ginelli}. Note that in the following we always consider covariant Lyapunov vectors even if we drop the term ``covariant''. Due to the equivalence of delay systems with conservative delay to systems with constant delay, which where extensively studied in the past and recent literature, we concentrate especially on the case of systems with dissipative delay. We are interested in structural properties of the Lyapunov vectors that are connected to the spectral properties of the Koopman operator following from the dynamics of the underlying access map. Hence the following analysis is intended to be independent of the dynamics of the delay system and concomitant questions like stablity and orbit structure. So let us again consider the simplest DDE given by Eq.~\eqref{eqn:lameDDE}.

\begin{figure}[htp]
\raggedright
\hspace{5mm} a)\\
\centering
\includegraphics[width=0.4\textwidth]{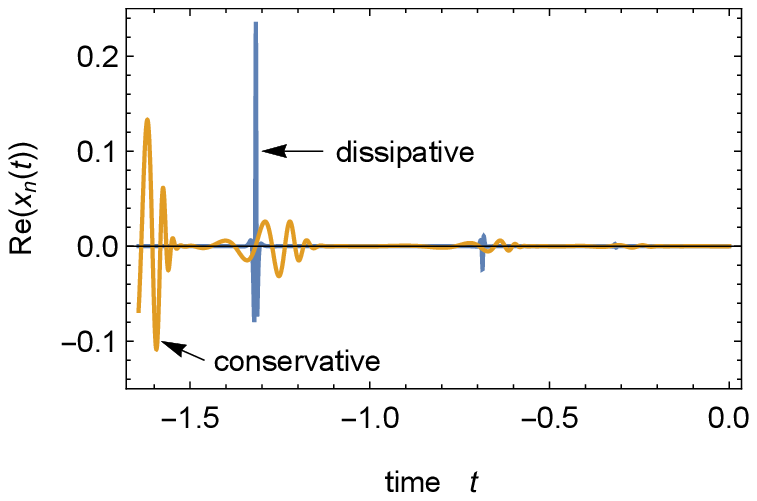}\\
\raggedright
\hspace{5mm} b)\\
\centering
\includegraphics[width=0.4\textwidth]{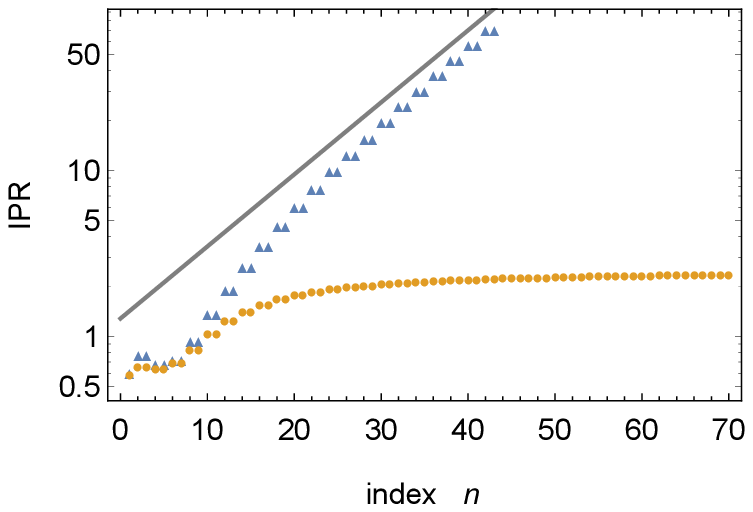}
\colorcaption{\label{fig:lyavecexample} a) Lyapunov vectors (real part), i.e., eigenvectors of the monodromy operator, and b) inverse participation ratio (IPR) of the periodic part $p_n(t)$ of the Floquet decomposition for exemplary conservative (dots) and dissipative (triangles) delays of Eq.~\eqref{eqn:lameDDE} with $\tau(t)=\tau_0+(0.9/2/\pi)\sin(2\pi\,t)$. For conservative delay the mean delay was set to $\tau_0=1.51$ and for dissipative delay $\tau_0=1.54$. The IPR is plotted versus the index of the corresponding Lyapunov exponent ordered from the largest to the smallest.}
\end{figure}

Note that in this simple case of a delay system with periodic delay the Lyapunov vectors are equal to the eigenvectors $x_n(t)=p_n(t)\,e^{(\lambda_n+\imath\omega_n)t}$ of the monodromy operator, where $p_n(t)$ is a periodic function with the same period as the delay's period $T$ and $\lambda_n$ equals the $n$-th Lyapunov exponent ordered from the largest to the smallest. In Fig.~\ref{fig:lyavecexample} some exemplary Lyapunov vectors for dissipative and conservative delay and the inverse participation ratio (IPR) of the periodic prefactors $p_n(t)$ given by \cite{thouless1974}
\begin{equation}
\text{IPR}(p_n(t))= \frac{\int_0^T dt\, |p_n(t)|^4}{\left( \int_0^T dt\, |p_n(t)|^2 \right)^2}
\end{equation}
related to the first Lyapunov vectors are shown. In Fig.~\ref{fig:lyavecexample} one can see that the exemplary Lyapunov vector corresponding to the case of dissipative delay shows sharper peaks than the vector corresponding to conservative delay. A qualitative difference can be noticed by comparing the IPR of the periodic part $p_n(t)$ for the two delay types. Obviously,  the IPR for dissipative delay increases much faster than the IPR for conservative delay. For dissipative delays the IPR seams to increase exponentially, which indicates that the peak width decreases exponentially with the index of the Lyapunov exponent and the corresponding Lyapunov vector. In this case the strongly localized peaks of the Lyapunov vectors are located at the members of the repulsive periodic orbits or at the repulsive fixed points.

In the following we use the aforementioned results for the Lyapunov vectors related to dissipative delay to develop a simple analytic model that describes the characteristic properties and gives us an idea how the structure of the Lyapunov vectors is connected to the scaling behavior of the Lyapunov spectrum and the related average divergence. So let us consider our simple delay system given by Eq.~\eqref{eqn:lameDDE}. We assume the reduced access map corresponding to the retarded argument $R(t)$ to possesses a repulsive fixed point at $t=0$ and attractive fixed points at $-1/2$ and $1/2$, i.e., the fixed points $t^*$ fulfill $R(t^*)=t^*-1$. Consequently, the initial function of our delay system is defined in the interval $[-1/2,1/2]$ whereto the time-intervals are mapped due to the Hale-Krasovkii notation, which was already mentioned in Sec.~\ref{sec:prelim}. Since this system is periodic, the solution operator mapping one state interval to the next state interval, which is given by the method steps, equals the monodromy operator. Hence, the Lyapunov vectors are equal to the eigenfunctions $x(t)$ of the monodromy operator and are solutions of the eigenvalue equation
\begin{equation}
\label{eqn:monodeig}
e^s\,x(t) = x(1/2) + \int_{-\frac{1}{2}}^{t} dt'\, x(\tilde{R}(t')), \quad t \in [-1/2,1/2],
\end{equation}
whereby the Lyapunov exponents are the real parts of the Floquet exponents $s$ and $\tilde{R}(t)=R(t)+1$. Since the Lyapunov vectors are localized at the members of the repulsive periodic orbits or fixed points, in our simple case they are localized at $t=0$. Hence the integration in Eq.~\eqref{eqn:monodeig} is dominated by the values of the integrand in a small neighborhood of $t=0$. Roughly speaking, we assume that the system ``sees'' a linear retarded argument $\tilde{R}(t) \approx b\,t$ for $t \approx 0$. Furthermore we assume the value of the Lyapunov vectors at the end of the state intervals to be negligible compared to the strongly localized peak around $t=0$ in the limit of small Lyapunov exponents, i.e. $x(1/2) \approx 0$, which implies $x(-1/2) \approx 0$ due to Floquet theory. Hence, we approximate Eq.~\eqref{eqn:monodeig} by
\begin{equation}
\label{eqn:monodeigapprox}
e^s\,x(t) = \int_{-\frac{1}{2}}^{t} dt'\, x(b\, t'), \quad t \in [-1/2,1/2],
\end{equation}
which denotes an eigenvalue equation of an operator acting on periodic functions. Since the corresponding eigenfunctions are expected to be strongly localized, it is reasonable to approximate their periodic extension by a sum of localized functions defined on the whole real line and shifted by integer multiples of the period 1. The Fourier transform of this periodic approximation is given by the product of a train of delta functions and an envelope, which is analog to the ``form factor'' of diffraction patterns of crystals in solid state physics related to the structure of the crystals base. Hence, the envelope of the Fourier transform of the approximate eigenfunctions connects the localized structure of the Lyapunov vectors to the eigenvalues $e^s$ and we simplify Eq.~\eqref{eqn:monodeigapprox} by its expansion to the whole real line ``reducing the crystal to one atom'', i.e.
\begin{equation}
\label{eqn:monodconv}
e^s\,x(t) = \int_{-\infty}^{t} dt'\, x(b\, t'),
\end{equation}
This expression can be interpreted as solution operator of the pantograph DDE $\dot{x}(t)=e^{-s} x(b\, t)$ corresponding to solutions that exist on the whole real line, whereby the equation is named by the mechanical part of a current collection system of an electric locomotive \cite{1971ockendon,*1971fox}. It is easy to see that this equation possesses solutions for arbitrary $s$. So the assumptions we made lead to the loss of the information that the Lyapunov spectrum is discrete, which is connected to the discreteness of the Floquet exponents for periodic delay equations. Not surprising, because we built our model to approximate the Lyapunov vectors of a delay system with any access map of any period that has a repulsive fixed point.

The right hand side of Eq.~\eqref{eqn:monodconv} can be expressed by a convolution using the Heaviside step function $\Theta(t)$. Hence we write
\begin{equation}
e^s\,x(t) = \Theta(t) * x(b\, t).
\end{equation}
Now its Fourier transform is trivially given by
\begin{equation}
\label{eqn:monodfouriert}
e^s\,X(f) = \frac{1}{2 b} \left( \delta(f) + \frac{1}{\imath \pi f} \right) X(f/b),
\end{equation}
where $X(f)$ denotes the Fourier transform of $x(t)$ with respect to the frequency $f$. For simplicity we assume the integral over the solutions of Eq.~\eqref{eqn:monodconv} to be equal to zero, i.e., we neglect the delta distribution in Eq.~\eqref{eqn:monodfouriert} and obtain after some rearrangements
\begin{equation}
\label{eqn:monodrescal}
\gamma \,X(f) = \frac{1}{f} X(f/b), \text{ where } \gamma := \imath 2\pi b\,e^s.
\end{equation}
With Eq.~\eqref{eqn:monodrescal} we have derived a linear first-order $q$-difference equation (cf. \cite{2012annaby}), which is solved by any linear combination of the solutions $X_1(f)$ and $X_2(f)$ given by
\begin{equation}
\label{eqn:solmonodrescal}
X_{1/2}(f)=\Theta(\pm f)\, e^{-\frac{1}{2\log b} \left( \log f + \log\gamma+\frac{\log b}{2} \right)^2}.
\end{equation}
We omit the derivation of the solution above, because it is a simple straightforward calculation by transforming the $q$-difference equation into a linear ordinary difference equation and does not contribute to a better understanding. 

\begin{figure}[htp]
\raggedright
\hspace{5mm} a)\\
\centering
\includegraphics[width=0.4\textwidth]{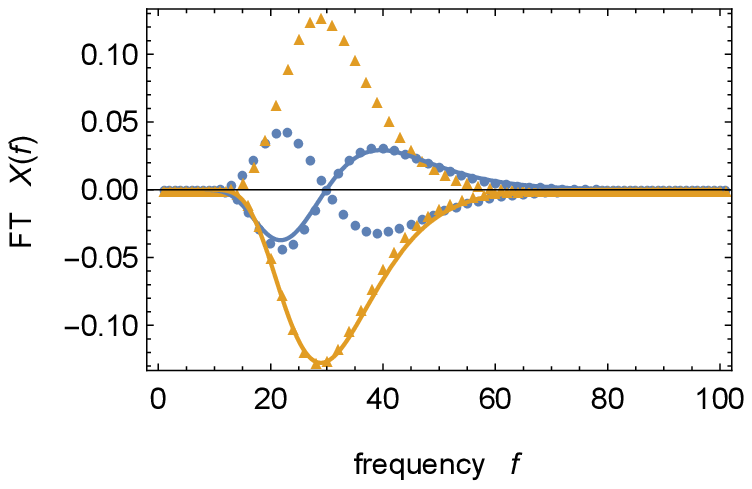}\\
\raggedright
\hspace{5mm} b)\\
\centering
\includegraphics[width=0.4\textwidth]{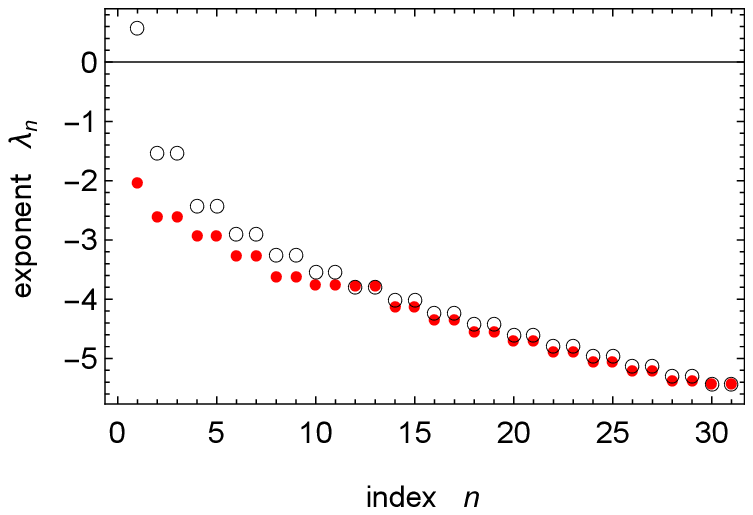}
\colorcaption{\label{fig:lyamod} a) Comparison of the real part (dots) and imaginary part (triangles) of the Fourier amplitudes of a Lyapunov vector (index $n=29$) with the corresponding envelope of the Fourier transform $X(f)\propto X_1(f)$ (lines), with $X_1(f)$ given by Eq.~\eqref{eqn:solmonodrescal}. b) Lyapunov spectrum from semi-discretization (empty circles) compared with modeled spectrum (dots) computed by fitting Eq.~\eqref{eqn:solmonodrescal} to the Fourier amplitudes of the Lyapunov vectors. The system Eq.~\eqref{eqn:lameDDE} with $R(t)=t-\frac{0.1}{2\pi} \sin (2\pi\,t)-1$ was used.}
\end{figure}

In Fig.~\ref{fig:lyamod}a our approximation of the envelope of the Fourier transform Eq.~\eqref{eqn:solmonodrescal} corresponding to a given small Lyapunov exponent of an exemplary system is compared to the Fourier amplitudes of the corresponding Lyapunov vector, whereby $b$ was set to the slope at the repulsive fixed point of the access map and $s$ was set to the numerically computed Floquet exponent. Additionally we computed the Lyapunov spectrum of the system by fitting a suitable linear combination of Eq.~\eqref{eqn:solmonodrescal} to the eigenfunctions of the monodromy operator by varying $\gamma$. The result is shown in Fig.~\ref{fig:lyamod}b. As desired, our simple model connects the structure of the Lyapunov vectors to the corresponding Lyapunov exponents and holds asymptotically for small Lyapunov exponents. Consequently our model indicates that the localization is caused by the repulsive fixed points or periodic orbits of the reduced access map and the spectral properties of the corresponding Koopman operator.

One interesting property of the Lyapunov vectors in the case of dissipative delay can be directly derived from Eq.~\eqref{eqn:monodrescal}. A solution $X(f)$ can be rescaled on the frequency domain to obtain other solutions. This means, if $X(f)$ is a solution with eigenvalue $\gamma$, $X(\beta f)$ is a solution with eigenvalue $\beta \gamma$ for arbitrary $\beta$. Consequently, the envelope of the Fourier transform of the Lyapunov vectors exponentially broadens with increasing index of the Lyapunov exponent and it follows that the Lyapunov vectors localize exponentially on the time domain as already shown in Fig.~\ref{fig:lyavecexample}b. This also has consequences for the finite section method introduced in Sec.~\ref{sec:finsec} and other algorithms for the computation of the Lyapunov spectrum. If the Lyapunov spectrum $\lambda_n$ asymptotically scales linear with the index $n$, the number of Fourier modes or similar basis vectors that are needed to obtain a good approximation of the $n$-th Lyapunov vector increases exponentially. Consequently, the number of well approximated Lyapunov exponents and Lyapunov vectors increases logarithmically with the dimension of the finite section approximation, which is worse for numerical analysis and also explains the slow convergence of the Lyapunov spectra for the systems with dissipative delay in Fig.~\ref{fig:spectra}b in Sec.~\ref{sec:class}. In other words the complexity of numerical algorithms increases exponentially. Moreover the question raises up whether the finite section approximations with respect to common orthonormal bases of the operators have anything to do with the exact operator, because the ratio between the number of well approximated exponents and  computed exponents or vectors tends to zero. Since the determinant is the product over all eigenvalues of a matrix, the determinant approach to compute the asymptotic scaling behavior of the Lyapunov spectrum will fail for typical orthonormal bases in the case of systems with dissipative delay. That is why we considered the Lyapunov base in Sec.~\ref{sec:specdelayclasses} for the numerical analysis.

\section{Conclusion}

In the present paper we have investigated the influence of the access map defined by the delay on the dynamics of delay systems with time-varying delay. In particular we have separated the dynamics due to the variable delay from the dynamics of the whole delay system by an operator theoretic approach, where the solution operator is decomposed into two different operators. Hence the time-evolution of delay systems with time-varying delay is characterized by the action of the Koopman operator corresponding to the access map and by the action of the constant delay evolution operator, which is similar to the solution operator of delay systems with constant delay. The decomposition of the dynamics into the deformation of the state by the Koopman operator and the integration by the constant delay solution operator connects the theory of DDEs with the theory of circle maps and the Koopman operator framework of the analysis of dynamical systems called ``Koopmanism''. This connection itself should have the potential to extend the view on the dynamics of systems with time-varying delay and can be a nice foundation for further research.

Using our abovementioned framework we have analyzed the influence of the two fundamental classes of delays on the Lyapunov spectrum and the evolution of small volumes on finite dimensional subspaces on the infinte dimensional state space of the delay system. The classification is determined by the question whether a system with time-varying delay is equivalent to a system with constant delay or not, which depends directly on the existence of a topological conjugacy between the access map and a pure rotation. We have shown that the asymptotic scaling behavior of the Lyapunov spectrum can be separated into a logarithmic part related to the constant delay evolution operator known from systems with constant delay and a part related to the Koopman operator depending on the dynamics of the underlying access map. If the access map is topological conjugate to a pure rotation, the delay system with time-varying delay is equivalent to a system with constant delay, the part of the Lyapunov spectrum related to the Koopman operator vanishes and we call the related delays conservative delay because the corresponding access map is a conservative system. Contrarily, if the conjugacy does not exist, the delay system is not equivalent to a system with constant delay and the asymptotic scaling behavior of the Lyapunov spectrum is dominated by the part related to the Koopman operator, which is linear with respect to the index of the Lyapunov exponent. Since the related delays lead to an additional dissipation by the Koopman operator, we call them dissipative delays. We obtain a similar dependence on the delay type for the evolution of small volumes on finite dimensional subspaces of the state space and their mean relaxation rate (average divergence). Due to the volume-preserving retarded access by the unitary Koopman operator, conservative delays lead to the same scaling behavior of the average divergence with respect to the dimension of the subspace as known from systems with constant delay, and dissipative delays lead to an additional contribution to the average divergence that describes the volume evolution under the action of the non-unitary Koopman operator.

Furthermore we have analyzed the influence of the delay type on the structure of the Lyapunov vectors and briefly investigated the consequences on numerical algorithms, particularly those including finite section approximations of the solution operator. The Lyapunov vectors of delay systems with dissipative delays localize exponentially, which leads to an exponential broadening of the envelope of the coefficients of their expansion in terms of common orthonormal bases. We have demonstrated that for dissipative delay the number of well approximated Lyapunov exponents or Floquet exponents raises only logarithmically with the dimension of the finite section approximation of the solution operator. So maybe it is worth to check whether this and similar algorithms can be improved by special bases, such as wavelet bases, which take into account the exponential localization of the Lyapunov vectors or eigenvectors of the monodromy operator in the case of periodic systems. Indeed, the inverse Fourier transform of the function appearing in Eq.~\eqref{eqn:solmonodrescal} can serve itself as a mother wavelet \cite{grossmann_transforms_1986}, and is nowadays also known as Log-Gabor wavelet. The consequences of this fact will be explored in a future publication.

\appendix

\section{\label{sec:applyaspec} Convergence of Lyapunov spectra}

\begin{figure}[h]
\raggedright
\hspace{5mm} a)\\
\centering
\includegraphics[width=0.4\textwidth]{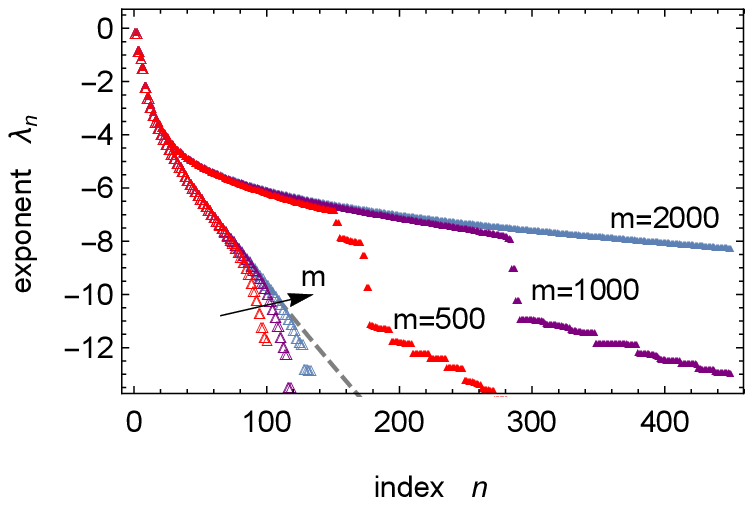}\\
\raggedright
\hspace{5mm} b)\\
\centering
\includegraphics[width=0.4\textwidth]{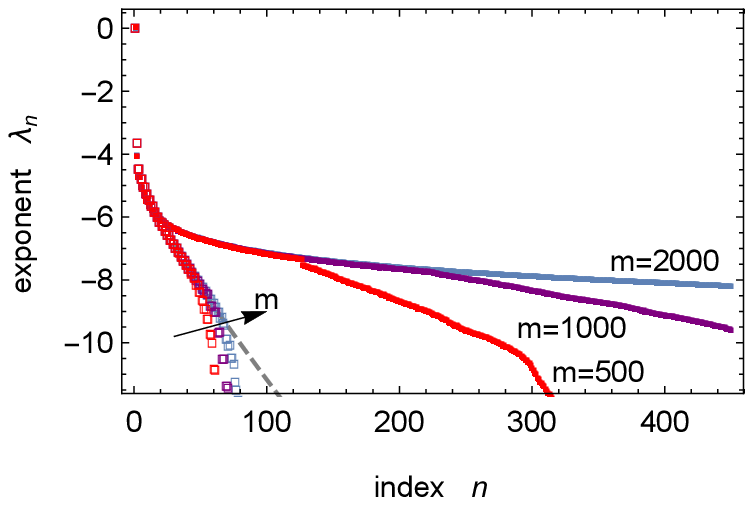}
\colorcaption{\label{fig:spectra_app} Comparison of the convergence properties of the Lyapunov spectra shown in Fig.~\ref{fig:spectra}a in Sec.~\ref{sec:class} between systems with conservative delay (filled) and dissipative delay (empty), computed by the \emph{semi-discretization} method \cite{2004insperger,2011insperger} for a) the turning model, given by Eq.~\eqref{eqn:drei}, (triangles)  and b) the Hutchinson equation, given by Eq.~\eqref{eqn:hutch} (squares). The parameter $m$ denotes the number of equidistant points approximating the state of the delay system. As a guide to the eye the asymptotic scaling behavior for dissipative delay, which is derived in Sec.~\ref{sec:specdelayclasses}, is represented by the dashed gray line.}
\end{figure}

In this section the convergence properties of the Lyapunov spectra of the turning model given by Eq.~\eqref{eqn:drei} and the Hutchinson equation given by Eq.~\eqref{eqn:hutch} are analyzed as done for the Mackey-Glass equation in Sec.~\ref{sec:class}. In Fig.~\ref{fig:spectra_app} the Lyapunov spectra of these systems for the two delay classes, conservative and dissipative delay, are shown under variation of the number $m$ of equidistant points approximating the state, whereby the same system parameters and delay parameters are used as in Sec.~\ref{sec:class}. As in the case of the Mackey-Glass equation, for dissipative delay the number of well approximated Lyapunov exponents increases much slower than for conservative delay. This is clearly a consequence of the localization properties of the Lyapunov vectors as derived in Sec.~\ref{sec:vec}.

\section{\label{sec:appmat} Matrix representations of the operators}

Let us consider the bases $\{\phi_{k i}(t)\}_{i\in\mathbb{N}}$ of the spaces $\mathcal{F}\left([t_{k-1},t_{k}],\mathbb{R}^d\right)$ and the corresponding dual sets $\{\psi_{k i}(t)\}_{i\in\mathbb{N}}$ in the dual spaces $\mathcal{F}^*$, as defined in Sec.~\ref{sec:finsec}. The mentioned sets define biorthogonal systems of our spaces $\mathcal{F}$ endowed with a weighted inner product such that
\begin{equation}
\int_{t_{k-1}}^{t_{k}} dt\, w_k(t)\, \psi_{k i}(t)\,\phi_{k j}(t) = \delta_{ij},
\end{equation}
where the $t_k$ are connected to the initial time $t_0$ by Eq.~\eqref{eqn:intbord} in Sec.~\ref{sec:prelim}. For a simpler computation of the matrix representation of operators between the spaces $\mathcal{F}\left([t_{k-1},t_{k}],\mathbb{R}^d\right)$ with different $k$ it will be convenient to define the bases $\{\phi_{k i}(t)\}_{i\in\mathbb{N}}$ and $\{\psi_{k i}(t)\}_{i\in\mathbb{N}}$ to be connected by a linear timescale transformation. Hence we set
\begin{eqnarray}
\phi_{(k+1) i}(t)&=& \phi_{k i}\left( h_k(t) \right)\\
\psi_{(k+1) i}(t)&=& \psi_{k i}\left( h_k(t) \right)\\
w_{k+1}(t)&=&\frac{t_k-t_{k-1}}{t_{k+1}-t_k} w_k\left( h_k(t) \right),
\end{eqnarray}
whereby
\begin{equation}
h_k(t)=\frac{t_k-t_{k-1}}{t_{k+1}-t_k} (t-t_k) + t_{k-1}
\end{equation}
maps $[t_k,t_{k+1}]$ to $[t_{k-1},t_k]$ by shifting and linear rescaling. This leads to the advantage that ``switching'' between the spaces by the linear transformation $h_k(t)$ is represented by the identity matrix, i.e.
\begin{equation}
\label{eqn:switchid}
\int_{t_{k}}^{t_{k+1}} dt\, w_{k+1}(t)\, \psi_{(k+1) i}(t)\,\phi_{k j}(h_k(t)) = \delta_{ij}.
\end{equation}
The state $\bm{x}_k(t)$ can be represented as an infinite-dimensional vector $\bm{q}_k$ composed of the $d$-dimensional coefficient vectors $\bm{q}_{k,i}$, which are given by
\begin{equation}
\bm{q}_{k,i} = \int_{t_{k-1}}^{t_{k}} dt\, w_{k}(t)\, \psi_{k i}(t)\,\bm{x}_k(t). 
\end{equation}
In the same way we represent the operators $\hat{I}_k$ and $\hat{C}_k$ by infinite-dimensional matrices composed of $(d\times d)$-matrices. Thus the submatrix $\left\{\bm{I}_k\right\}_{ij}$ of the matrix representation of the constant delay evolution operator $\hat{I}_k$ is given by
\begin{equation}
\label{eqn:CDEOmat}
\left\{\bm{I}_k\right\}_{ij}= \int_{t_{k}}^{t_{k+1}} dt\, w_{k+1}(t)\, \psi_{(k+1) i}(t) \, \left( \hat{I}_k \, \phi_{(k+1) j}(t') \right), 
\end{equation}
whereby each component of the matrix valued operator $\hat{I}_k$ is applied to the scalar basis function $\phi_{(k+1) j}(t)$ and each component of the result is projected subsequently to the initial base by the weighted scalar product with $\psi_{(k+1) i}(t)$. The submatrix $\left\{\bm{C}_k\right\}_{ij}$ of the matrix representation of the access operator $\hat{C}_k$ is computed by 
\begin{equation}
\label{eqn:cproj}
\left\{\bm{C}_k\right\}_{ij}=\int_{t_{k}}^{t_{k+1}} dt\, w_{k+1}(t)\, \psi_{(k+1) i}(t) \, \left( \hat{C}_k \, \phi_{k j}(t') \right),
\end{equation}
where the computation is done as above with the difference that $\hat{C}_k$ changes the underlying space of the state. Hence we utilize the orthogonality relation Eq.~\eqref{eqn:switchid}.


\bibliography{ddespec}

\end{document}